%% file: main.tex
\documentclass[manuscript,screen]{acmart}
\usepackage{graphicx}
\usepackage{float} 
\usepackage{subfigure}
\usepackage{tabularx}
\usepackage{multirow}
\usepackage{color}
\usepackage{diagbox}
\usepackage{makecell}
\AtBeginDocument{%
  \providecommand\BibTeX{{%
    \normalfont B\kern-0.5em{\scshape i\kern-0.25em b}\kern-0.8em\TeX}}}

\acmConference[Proc. ACM Interact. Mob. Wearable Ubiquitous Technol.]{Make sure to enter the correct
  conference title from your rights confirmation email}{2024}{NY}
\acmPrice{00}
\acmISBN{978-1-4503-XXXX-X/18/06}

\begin{document}

\title{What Impacts the Quality of the User Answers when Asked about the Current Context?}

\renewcommand{\shorttitle}{What Impacts the Quality of the User Answers}


\author{Ivano Bison}
\affiliation{%
 \institution{University of Trento}
 \streetaddress{via Sommarive, 9}
 \city{Trento}
 \country{Italy}}
\email{Ivano Bison@unitn.it}

\author{Haonan Zhao}
\affiliation{%
  \institution{University of Trento}
  \streetaddress{via Sommarive, 9}
  \city{Trento}
  \country{Italy}}
\email{Haonan Zhao@unitn.it}

\author{Fausto Giunchiglia}
\affiliation{%
  \institution{University of Trento}
  \streetaddress{via Sommarive, 9}
  \city{Trento}
  \country{Italy}}
\email{Fausto Giunchiglia@unitn.it}

\renewcommand{\shortauthors}{I. Bison et al.}


\begin{abstract}
Sensor data provide an \textit{objective} view of reality but fail to capture the \textit{subjective} motivations behind an individual's behavior. This latter information is crucial for learning about the various dimensions of the personal context, thus increasing predictability. The main limitation is the human input, which is often not of the \textit{quality} that is needed. The work so far has focused on the usually high number of missing answers. The focus of this paper is on  \textit{the number of mistakes} made when answering  questions. Three are the main contributions of this paper. First, we show that the user's \textit{reaction time}, i.e., the time before starting to respond, is the main cause of a low answer quality, where its effects are both direct and indirect, the latter relating to its impact on the \textit{completion time}, i.e., the time taken to compile the response.
Second, we identify the specific \textit{exogenous} (e.g., the situational or temporal context) and \textit{endogenous} (e.g., mood, personality traits) factors which have an influence on the reaction time, as well as on the completion time.  Third, we show how reaction and completion time compose their effects on the answer quality. The paper concludes with a set of actionable recommendations.

\end{abstract}



\maketitle

\input{section/1.Introduction}

\input{section/2.Related_Work}

\input{section/3.Method}

\section{Analysis}
\label{sec:new}

In the analysis reported below, we use the data from the 158 respondents who have filled out the notifications for all 14 days of the first two weeks, that have provided at least 300 valid answers, and that have agreed to share their GPS location.  We have used the data only from the first two weeks based on the guidelines from \cite{stone2002capturing}. The main motivation is that, because of the different types of data collection, given the analysis performed in this article, the datasets from the first and the second two weeks are not comparable.  
In the first two weeks, a total of 58340 observations were collected,where we excluded sleep-related events, and GPS values where radius accuracy was greater than 100 meters. The analysis for Q3 is based on a sub-sample of 78 respondents and 7507 events. This reduction in number is motivated by the lack of information about the precise location of the relatives' home; this has forced us to exclude all respondents who, during the observation window, moved from the university to the parents' home and vice versa. (Both home and the relatives' home are possible answers in Fig. \ref{Time diary}.)

In the analysis of Q1, the dependent variable modeling the reaction time is the \textit{time}, measured in minutes, \textit{elapsed between the server sending the notification to the device and the moment the user starts answering}. To limit the effects of memory errors, all reaction times exceeding 8 hours are not considered. Similarly, technical problems, e.g., noise in an answer, due to transmission errors or when the reception time exceeds 150 minutes are also excluded.  
In the analysis of Q2, the dependent variable modeling the completion time is based on the time taken to complete the four questions in Fig.\ref{Time diary}. In the analysis we do not consider filling times of more than 75 seconds. 
In the analysis of Q3, the reaction time and completion time are those computed by the models for Q1 and Q2 respectively. No cases are dropped, and the available cases are analyzed using a linear regression model. Sleep events, compilation times of more than 75 seconds and notifications of more than 8 hours of age are excluded from the analysis. Furthermore, to reduce the uncertainty, we have considered only GPS positions with an accuracy of fewer than 100 meters. This has also allowed us to exclude other errors, for instance, confusing the parents' house with the student house, where we have considered only distances of less than 4000 meters. Given the above assumptions, the total number of notifications we have analyzed is 7507. 

Finally, the students' activities and locations, as from Fig. \ref{Time diary}, have been reclassified as follows: 
\begin{itemize}
    \item \textbf{Activities}: (1) Personal care; (2) Eating; (3) Study alone or with others; (4) Classroom lecture; (5) Social life \& Break: Social life, Coffee break; (6) Watching YouTube/Tv-shows, etc.; (7) social media/Phone/Chat: Facebook, Instagram, etc., on the phone/chat; (8) Free time: Reading a book; listen to music; Movie, Theater, Concert, Exhibit etc.; Shopping; Sport; Rest/nap; Hobbies; (9) Work: Housework, Work, Other activities; (10) Travel: By car, By foot, By bike, By bus, By train, By motorbike; 
    \item \textbf{Place}: (1) Home, Apartment, Room; (2) Relatives (house); (3) House (friends others); (4) University: Classroom/Lab., Classroom/Study Hall, Library; Other university place, Canteen; (5) Shop/Pub/Theatre: Shop supermarket etc., Pizzeria, pub, bar, restaurant, Movie Theater, Museum, etc.; (6) Workplace; (7) Other place: Other Library; Gym; Other place; (8) Outdoors; (9) Moving.
\end{itemize}
\noindent
The further data considered in the analysis are
\begin{itemize}
    \item the \textit{time context}, namely the specific moment and time when a question or an answer occurred, as collected by APPX.
\item the \textit{computing context}  was automatically collected using the smartphone internal hardware (e.g., GPS, accelerometer, gyroscope) as well as
the data collected by the so-called software sensors, e.g., the applications running on the device, see Fig. \ref{sensor data}.
\end{itemize}
\noindent
Based on the assumptions described above,
Section \ref{H1} reports the analysis for Q1, Section \ref{H2} reports the analysis for Q2, while Section \ref{sec:H3} the final and conclusive analysis for Q3, which builds on top of the results of the first two subsections.

\input{section/4.Reaction_time}

\input{section/5.Completion_time}

\input{section/6.Answer_correctness}

\input{section/7.Description}

\input{section/8.Conclusion}
\bibliographystyle{ACM-Reference-Format}
\bibliography{bibliography,new}

\end{document}

%% file: section/1.Introduction.tex
\section{Introduction}
\label{sec:intro}

Various studies have highlighted how predictable various aspects of human behavior are, see, for instance, the work on mobility \cite{brockmann2006scaling,gonzalez2008understanding,song2010limits,cuttone2018understanding,alessandretti2020scales}, social interactions \cite{eagle2009eigenbehaviors,eagle2009inferring}, or people preferences for their favorite places \cite{alessandretti2018evidence} and friends \cite{miritello2013limited}. Some of these studies show that contextual information is useful in applications such as health and physical activity monitoring \cite{rabbi2015mybehavior,intille2016precision,yue2020deep,yue2021exploring}, mental health monitoring \cite{wang2018tracking,wang2020social,yue2019mimic} or elderly care \cite{lee2015sensor,berke2011objective,wang2021self,yue2021intention}, and also for predicting the individuals' behaviours and traits \cite{do2012contextual,harari2016using,wang2018sensing,peltonen2020phones}. In this latter case, the challenge is how to compute a high-quality characterization of this type of information \cite{lane2010survey}. In this line of thought, the work in \cite{bettini2010survey,riboni2011owl,helaoui2013probabilistic,suchman1987plans,holtzblatt1997contextual} and in 
\cite{radu2018multimodal,vaizman2017recognizing,vaizman2018context,bradley2005toward} are, respectively, examples of early and more recent work on the general topic of \textit{context recognition}.
Nevertheless, most such studies have concentrated on the use of (only)  \textit{sensor} data. This, in turn, has generated various kinds of errors, most noticeably \textit{data validity}, namely the accuracy of the indicator of the phenomenon being measured, and \textit{data completeness}, where the occurrence of missing data should be random rather than systematic. 

The main limitation is that, while providing a good \textit{objective} representation of reality, sensors are unaware of the \textit{subjective} motivations behind a given individual's activities.
Starting from the consideration that human behaviour is based on the individual's subjective perception of the current context, a few studies have put the role of context at the core of the analysis \cite{zhang2021putting}. These studies use sensor data as additional, even if crucial, information. They show that
when a user-provided subjective description of the current context is available,
any target modality (e.g., where the person is, or what s/he does) becomes substantially more predictable if one also exploits information about the other modalities (e.g., time, user characteristics, social ties). This opens the possibility for the machine to collect information and learn about every aspect of the daily life of a person, with high-impact applications in all research areas focusing on the flow of individuals' behaviour and thinking \cite{hormuth1986sampling, wilhelm2012conducting} and the ecology of human development \cite{bronfenbrenner1977toward}. Examples of applications are in Medical behaviour \cite{stone2007science}, Clinical Psychology, Social Sciences, Human-Computer Interaction and, lately, Human-in-the-loop Artificial Intelligence (AI) \cite{KD-2022-Bontempelli-lifelong}.
But, for this information to become available, there is a need of an active collaboration of the user with the machine, where this collaboration has a main limitation in that the human input is often not accurate, see, e.g., \cite{H-2016-Huang,H-2014-Wang}. 

The goal of this paper is to provide an in-depth study of which \textit{factors} influence the \textit{quality} of the answers that users provide when asked \textit{in the wild}. Here by quality, we mean \textit{a low number of errors} in the answers themselves. We focus on two main types of \textit{factors}, that we call \textit{exogenous} and \textit{endogenous}, which
influence the overall behaviour of an individual and, in turn, the answer quality. Examples of exogenous factors are the physical and social \textit{situational context} (e.g., where users are, what they are doing, who they are with), the \textit{temporal context} (e.g., the day of the week), and the \textit{computing context}, (e.g., network connectivity, communication bandwidth). Examples of endogenous factors are the user's \textit{personality}, \textit{cognitive and emotional} states (e.g., mood, burden).
The analysis provided follows the Experience Sampling Methodology (ESM), see, e.g., \cite{S-2014-Larson,stone2007science,zeni2021improving}, where the reference dataset has been built via an interval-based data collection from 158 University students over a period of two weeks, including 58,340 answers with corresponding GPS positions \cite{KD-2021-bison-SU2} \footnote{A clean version of this dataset, GDPR compliant and suitably anonymized, can be downloaded from the \texttt{LivePeople} catalog at the URL
\url{https://datascientiafoundation.github.io/LivePeople}. The search keyword \texttt{SU2} returns all the datasets associated to this data collection. \texttt{LivePeople} contains extensive information about the datasets if indexes, including:
(a)  A technical report describing the details of the data collection; (b) The dataset documentation and metadata
and (c) the procedure to be followed in order to request the dataset. To be fully compliant with \texttt{GDPR}, a licence must be signed before downloading the dataset.}.
This is in line with both recent ESM studies \cite{van2017experience} where the decision of a duration of 2-4 weeks duration follows the recommendation by Stone et al. \cite{stone1991measuring}. This experiment was crucially based on the use of an EMA/ESM application named iLog \cite{KD-2014-PERCOM,zhao2024human}, which allows to collect sensor data, typically but not only, from smartphones, and to ask questions about the user's situational context \cite{KD-2017-PERCOM,KD-1993-giunchiglia}, in the form of \textit{time diaries}, i.e., sets of questions asked multiple times, in various programmable moments of the day \cite{S-1939-Bowers}
\footnote{The version of time surveys used in this paper is based on \textit{HETUS} (\textit{Harmonised European Time Use Surveys}), see the details at the URL \url{https://ec.europa.eu/eurostat/web/microdata/harmonised-european-time-use-surveys}.}.

The analysis described in this paper provides the following insights: 

\begin{enumerate}
 
    \item The elapsed time from notification to answer, i.e., the \textit{reaction time}, relates to the situational and temporal context, as well as to mood and procrastination. The behaviour also changes during the observation period where, over time, users seem to find their own routine inn the usage of iLog.
    
    \item The time to fill in a question, i.e., the \textit{completion time}, relates to delays in the reaction time, the cognitive ability of the respondent and also possible disturbance or multitasking effects due to the social context.
    
    \item The \textit{quality of an answer} is influenced by both the reaction and the completion time. However the main impact is from the reaction time which, as from the previous item, has also an effect on the completion time. 

       \item The number of wrong answers is substantial. In the test case we used in the experiment there were 730 answering errors out of a total number of 7624 questions, for an overall percentage of 9.6\% of wrong answers. This percentage looks even more relevant if we take into account the specific question we tested, that is, whether the user was at home, a type of question that  would seem impossible to get wrong.
    
    \item Two main lessons about how to improve the quality of answers in future experiments where the user is asked to answer questions about the current context. The first is that \textit{reaction time} is \textit{the main variable to keep in control}. The second is that the most effective to control reaction time is \textit{by operating on the various types of context}, as detailed below, being these more controllable than endogenous factors, e.g., procrastination, personality or mood.
\end{enumerate}

\noindent
The paper is organized as follows. In Section \ref{sec:related} we provide the related work. In Section \ref{sec:method} we describe the overall methodology. In Section \ref{sec:new} we describe the analysis on reaction time, completion time and answer quality, respectively. In Section \ref{sec:des} we discuss and highlight the main take-away lessons which result from this work, including a set of recommendations for future ESM studies (Section 5.3). Finally, Section \ref{sec:conclusion} concludes the paper.

%% file: section/2.Related_Work.tex
\section{Related Work}
\label{sec:related}

Mobile self-reports are a popular technique for collecting data from participants in the wild. Because of its potential ability to record participants' behaviour, this type of data allows for the development of a ground-truth, where the meaning of the data itself is provided directly by the user. 
One of the main problems is the impossibility of capturing the real causes of mistakes, mainly because of the impossibility of observing the behaviour of the respondent \textit{in-the-wild} (e.g., which causes? which conditions?) \cite{S-2019-Wenz}. A further complications is that researchers have no or  little control on the surrounding environment. The user cannot be "supervised" by a human interviewer who can decide when and where s/he should answer the query.  
Despite all of this, the assessment of the accuracy of responses has received little attention in the literature. Researchers tend to wrongly assume that responses are accurate, without further validating this assumption \cite{H-2019-Van}.  

The problem of low-quality responses has been extensively studied in the EMA/ESM research \cite{shiffman2008ecological,S-2014-Larson}, where the results have then been incorporated in more complex and sophisticated research protocols \cite{S-2014-Larson,csikszentmihalyi1992experience,hektner2007experience}. However, the literature has mainly focused on increasing the \textit{response rate} of participants. For example, Boukhechba et al. \cite{H-2018-Boukhechba} found a higher response when ESM questions are sent after phone calls, with respect to social media usage. Similarly, Berkel et al. \cite{H-2020-Van} found that active phone use before an ESM question could increase the response rate. Looking at the context, some work has focused on how location \cite{H-2021-Sun}, activities \cite{H-2017-Mishra}, different times during the day \cite{H-2017-Pielot}, and social interactions \cite{H-2018-Boukhechba} impact the response rate. Berkel et al. \cite{H-2019-Van} concentrated on the problem of correctness and found that the highest accuracy values are obtained when the participant' screen is turned off at notification arrival (that is, when the phone is not being used). They also show that longer completion times of questions generate answers with lower accuracy; the main limitation of this work is that the only contextual factor considered is the smartphone usage. 

In Artificial Intelligence, related work has been done on understanding the users' reactions to notifications. Here the focus is on the \textit{reaction time} and on how to predict the probability of a user checking a notification on time. Iqbal et al. \cite{H-2010-Iqbal} show that users more easily accept disruptions from messages carrying useful information. Fischer et al. \cite{fischer2010effects} found that content is more important than notification time. The participants are willing to be interrupted by what they are interested in. Moreover, the same authors \cite{H-2011-Fischer} have investigated the dependence of the response time on the previous types of interaction (e.g., a phone call or an SMS). Ho and Intille \cite{H-2015-Oh, ho2005using} developed an application based on the idea that a transition between two physical activities (e.g., sitting and standing) is a suitable moment for a notification (i.e., phone calls, messages and reminders). In another study, Mehrotra et al. \cite{H-2015-Mehrotra.b} suggested using context information and application usage in order to predict the best notification time. Aminikhanghahi et al. \cite{aminikhanghahi2019context} worked on combining the ESM technology with knowledge of the user activities. 

The \textit{Total Survey Error paradigm}, as elaborated in Sociology, provides a theoretical framework for optimizing surveys  by maximizing data quality within budgetary constraints \cite{sen2019total,amaya2020total}. According to this framework  there are four main sources of error which are independent among one another while correlating with the phenomena under study. These factors can be summarised as follows:
\begin{enumerate}
    \item  The \textit{situational and temporal context} in which the user inputs information into the smartphone \cite{S-2007-Lavrakas,force2010new,S-2019-Wenz}. 
    \item The \textit{cognitive task} involved in the response process \cite{ weisberg2009total}, in time-related questions in the multi-component approaches \cite{S-2013-Lynn} as well as in respondent motivation two-track theories \cite{S-1981-Cannell,S-1991-Krosnick,S-1987-Krosnick}. 
    \item Those \textit{conscious or unconscious factors,} e.g., personality, attitudes and habits, that influence the user's behaviour \cite{S-2013-Lynn,S-2019-Read}.
    \item The technical problems related to the functioning of the  technology, e.g., phone and phone app \cite{grammenos2018you,schilit1994context,fielding2016sage,sarmadi2023review,balto2016accuracy,struminskaya2020augmenting}. 
\end{enumerate}
These four sources are assumed to generate errors 
with \textit{any} question-answering process and, therefore, also with mobility self-reports. This hypothesis is the main theoretical foundation on which the work presented in this paper is based. As far as we know, this hypothesis has never being applied to mobility self-reports, in particular in the scenario of asking questions about context. 
In this perspective, the work in this paper can be seen as validating and detailing the specific factors of how the hypotheses of the \textit{Total Survey Error paradigm}  get instantiated in this scenario.

%% file: section/3.Method.tex
\section{Methodology}
\label{sec:method}

We organize this section as follows. In Section \ref{sec-rq} we articulate the motivations behind the selection of the model and, consequently, the three research questions Q1, Q2 and Q3.  In Section \ref{sec-33} we describe the sample selection process and the scheduling of the time diaries.
Finally, in Section \ref{sec-32}, we synthetically describe the three statistical models used for the analysis of Q1, Q2, and Q3, respectively.

\subsection{The overall model}
\label{sec-rq}

\begin{figure}[htp]
\centering
\includegraphics[scale=0.65]{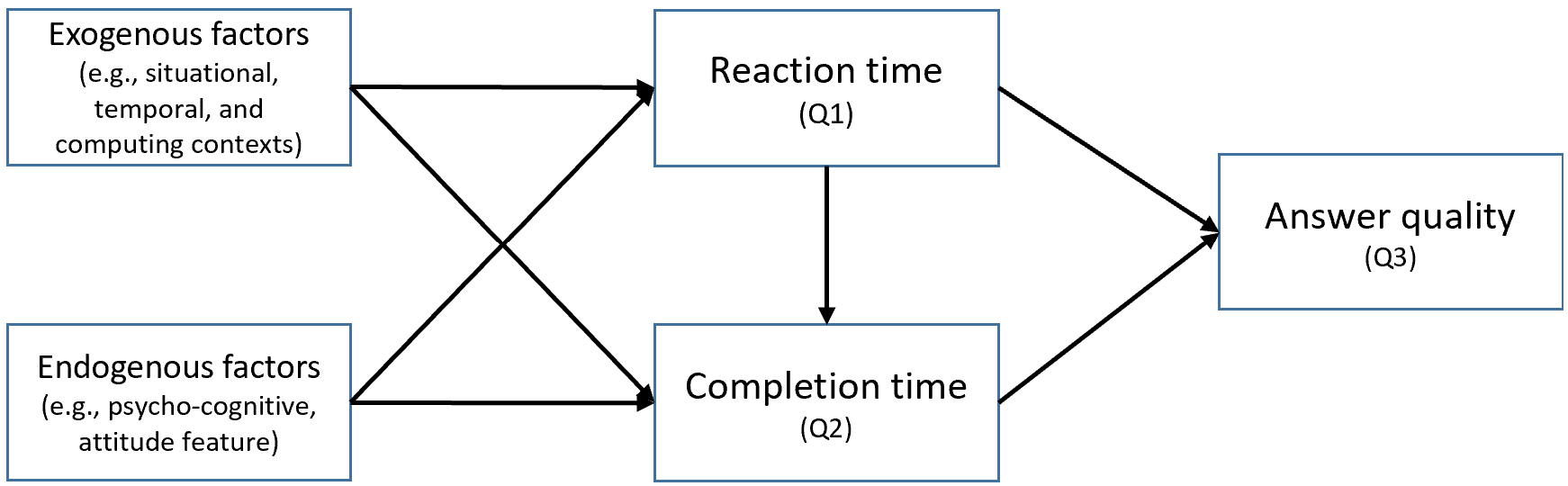}
\caption{The causal chain of events impacting the answer quality.}
\label{ The theoretical model}
\end{figure}

The key issue is to identify the key factors which influence the quality of an answer, where these factors must be operationalizable as part of a strategy which allows a machine to get the best possible results out of the user answers. The starting point is the work from the \textit{Total Survey Error paradigm} (see Section \ref{sec:related}). Based on this work, we assume
 the existence of a causal chain of events which influences the quality of answers and that we asbtractly model according to the schema sketched in Fig. \ref{ The theoretical model}. 

\noindent
Proceeding from left to right we distinguish between \textit{endogenous} and \textit{exogenous factors},
that play a role and have an impact on the quality of answers. These two sources of error can be detailed as follows.  
\begin{itemize}
\item \textit{exogenous factors}, mainly related to the context of use, that we organize along three dimensions, that is: the \textit{physical } and  \textit{social situational context}, the \textit{temporal context} and the  \textit{computing context}. The first two dimensions relate to factors such as the degree and the type of distractions, the presence of others, or the multitasking behaviour – whether on the same device, on a different device, or even on a different medium \cite{S-2007-Lavrakas,S-2010-Lavrakas,S-2013-Lynn,S-2014-Link,S-2019-Read,S-2019-Wenz}. The third dimension, i.e., the computing context, relates to the technical problems connected to the functioning of iLog, taking into account of all the possible things which may go wrong \cite{heron2017using}. 
    
\item \textit{endogenous factors}, that we organize along two main dimensions. The first is the user \textit{characteristics and behaviour}, for instance, the familiarity or comfort with the device, how the respondent uses the device (i.e., frequencies of use, duration and frequency of browsing session) \cite{S-2014-Link,S-2011-Groves,S-2009-Lai,S-2009-Raento,S-2013-Lynn,S-2019-Read,S-2017-Jackle}. The second is the \textit{willingness and ability} to follow and complete a task on a device \cite{S-2011-Groves,S-2013-Lynn,S-2005-Oulasvirta}. Relevant to this factor is a series of psycho-physical attitudes, personality, behaviour and habits, conscious or unconscious, that influence the individual \cite{S-2013-Lynn,S-2019-Read}. These cognitive performances differ across individuals \cite{cowan2012working}, where human cognition is also time-variant. Notice that the person mental state \cite{beilock2007poor,cowan2010magical}, as well as the some  exogenous factors, may influence cognitive performance. Examples of these exogenous factors are the time of the day \cite{schmidt2007time,west2002effects},  and smartphone usage \cite{hyman2010did,kushlev2016silence}. 

\end{itemize}
We organize our analysis around three research questions, Q1, Q2, Q3, which allow to model the chain of effects of how exogenous and endogenous factors influence the quality of answers. We have the following.
\begin{itemize}
    \item  \textit{Reaction time} (Q1). The goal of Q1 is to understand \textit{which endogenous as well as exogenous factors have an impact on the reaction time}, where the reaction time is defined as the time between receiving a notification and initiating a response. 
    \item \textit{Completion time} (Q2). The goal of Q2 is to understand \textit{which endogenous as well as exogenous factors have an impact on the  completion time}, where the completion time is defined as the time it takes to complete a response, starting from the end of the reaction time.
    \item \textit{Answer quality} (Q3). The goal of Q3 is to assess and validate the causal chain, as represented in Fig.\ref{ The theoretical model}, by providing a quantitative evaluation of \textit{how reaction and completion times jointly affect the answer quality.} 
\end{itemize}
\noindent
Let us focus on Q1, Q2 and Q3.  

\subsubsection{Q1: The reaction time}
The optimal user behaviour in data collection would be that the user responds as soon as s/he receives the notification. The reason is that, the longer the elapsed time, the higher is the risk of a memory error (mainly related to forgetting) and, consequently, a higher probability of a wrong answer \cite{mccabe2012guide}. 

\begin{figure}[htp]
\centering
\includegraphics[scale=0.45]{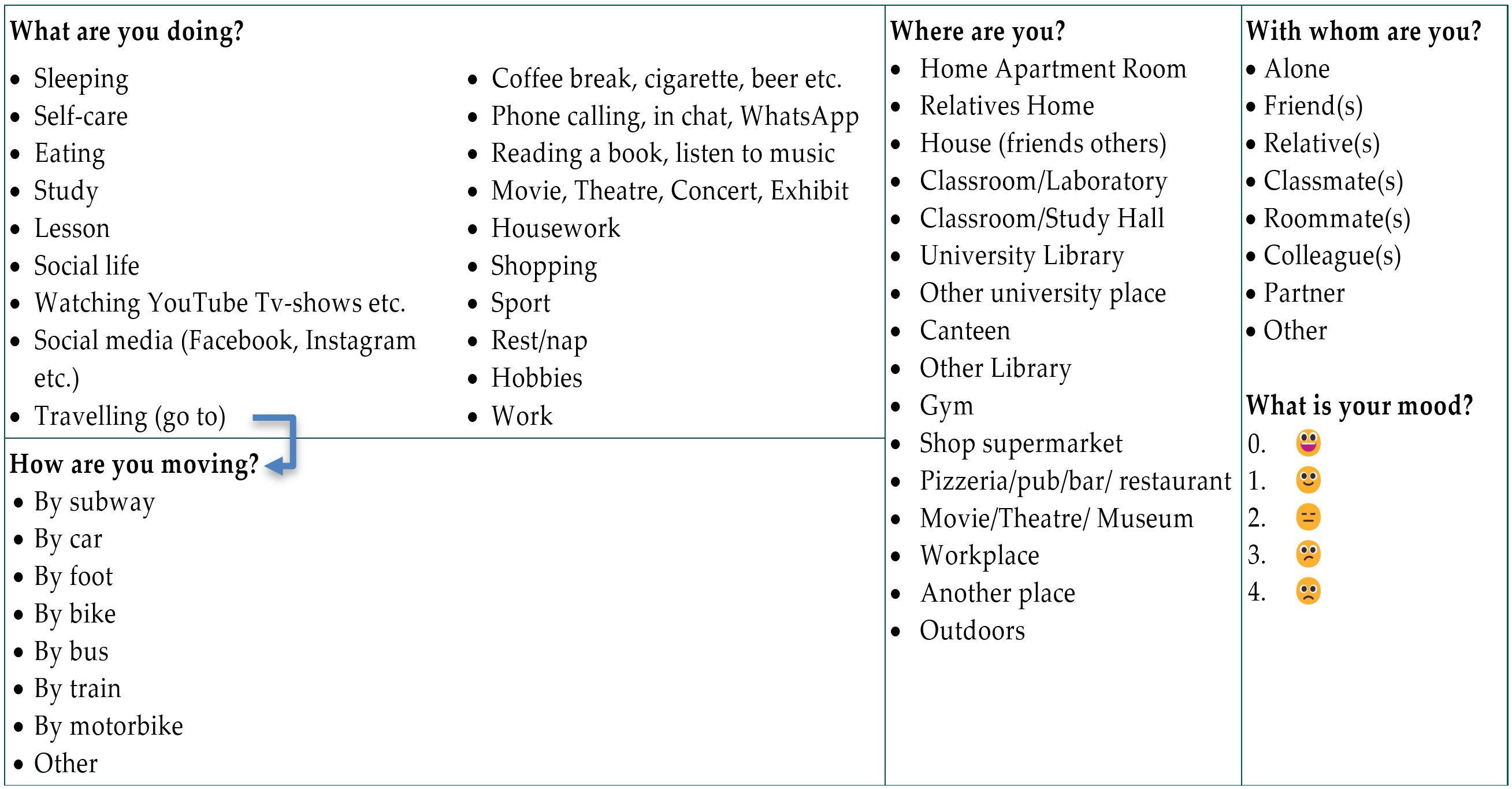}
\caption{Time diary.}
\label{Time diary}
\end{figure}

We focus our attention on three factors which may have an impact on the reaction time, as follows:

\begin{itemize}
    
\item \textit{context history} \cite{chen2000survey} (an exogenous factor), namely a mix of user context (the social situation), physical context, computing context (network connectivity, etc.) and time context. The \textit{social and the physical context} are taken into account with three multiple choices self-reported time diary information: (a) "\textbf{What are you doing?}" captures the distracting effects of the activities the student is doing at the time; (b) "\textbf{Where are you?}" captures the environmental context and the consequent distracting effects due to where s/he is at that moment; (c) "\textbf{With whom are you?}" captures the disturbing effects of the social context. The details of these three questions and of the possible answers is reported in Fig.\ref{Time diary}. The question "\textbf{How are you moving?}" is asked any time the user answer that s/he is travelling.
The \textit{computing context} is considered as the \textbf{time elapsed}, in seconds, from when the notification is sent from the server to when the smartphone receives it. Finally, the \textit{time context} accounts for how different weekly activities can affect the reaction time. In the model, this latter idea has been operationalized as a distinction between weekdays and weekends. 

\item \textit{motivation and burden} (an endogenous factor) concerning both the effort (time and resources) and the degree of difficulty. In the model, this factor is modeled as a quadratic function of the \textbf{day of study}, quantified in a range from 0 (first day in the experiment) to 13 (fourteenth day in the experiment). 

\item  \textit{user characteristics} (an endogenous factor), modeled in terms of psychological traits and emotional status over the day. In fact, mood states and personality traits play a crucial role in the processing of emotion-congruent information across different cognitive tasks \cite{rusting1998personality}. 
This factor was taken into account by asking,  the user about his/her  \textbf{procrastination} syndrome and \textbf{emotional mood state}. The first question was done using the \textit{Irrational Procrastination Scale (IPS)}  \cite{S-2007-Steel,S-2010-Steel}. 
The second question was organized as a 5-point Likert scale, with options ranging from happy (0) to sad (4) \cite{LORR198937,killgore1999visual}, see Fig.\ref{Time diary}.
\end{itemize}

\subsubsection{Q2: The completion time}

As from above, a delayed answer may be the cause of a memory error. This applies to the reaction time but also to the completion time.
In particular, the cognitive process enabling the response processing may generate a long completion time and, consequently, have an impact on the response accuracy \cite{mccabe2012guide}. 
Furthermore, according to the multi-component approaches \cite{S-2000-Tourangeau}, the respondent's ability to focus on something specific while ignoring other stimuli \cite{S-2015-Kellogg} has an impact on the quality of an answer; also in this case, the longer the completion time, the lower the expected answer quality. 
This seems to suggest that shorter completion times generate a higher answer quality, as they decrease the risk of memory errors \cite{sudman1973effects}. However, this is not necessarily the case. First of all, with a fast completion time the probability of inaccuracy or typing errors is high \cite{S-2015-Callegaro,mccabe2012guide}. Furthermore, as from \cite{S-1984-Tourangeau,S-2000-Tourangeau}, the answer completion process is organized in four steps as follows: (1) comprehension of the question; (2) retrieval of relevant information from memory; (3) judgment required by the question; and (4) selection of an answer. An increase in the average compilation time may mean more time spent retrieving relevant information from memory, thus having a positive effect on the quality of an answer. While the negative effect of an increase of the reaction time seems to be established, the same cannot be said in the case of the completion time.

In the following we consider the following four factors as the possible sources of changes in the completion time.

\begin{itemize}
    \item  the \textit{competence in the usage of iLog}. This competence grows in time, and the result is a rapid reduction of the completion time which is quite rapid even in the early days. We model this learning process using two components. The first is \textit{the date/time of use}, measured as the \textbf{day of study}. The second is the \textit{memory of the response alternatives} as they physically appear \textit{in the list of answers}. The more a response modality is used, the more likely a respondent will remember its exact location and the faster the user will answer. We model the  \textbf{list of activities}  in the time diary as a pseudo-continuous variable where each activity is associated with its percentage of appearance. Thus, for instance, the activity ``studying" is replaced with its occurrence percentage 19.04\%, ``eating" with 4.61\%, and so on \cite{chen2014assignment}. 
    
\item
the \textit{reaction time}. One or more notifications may remain unanswered, so as to form blocks of notifications that the user can quickly fill after one another in a single response session. We model this factor as \textbf{the total number of pending notifications} which exists when the respondent starts the completion process.

\item the \textit{social context}. This factor models the disturbance or multitasking effects due to the presence of other people during the completion process \cite{S-2015-Kellogg}. We model this factor by taking into account whether the respondent is \textbf{alone} at the time he has to respond (see Fig.\ref{Time diary}).

\item the \textit{psycho-social} aspects. According to the literature, both \textbf{mood} and \textbf{procrastination syndrome} have an important role on memory and motivation \cite{S-2013-Forgas,S-2007-Steel} and, in turn, on the completion process.

\end{itemize}

\begin{figure}[htp]
\centering
\includegraphics[scale=0.55]{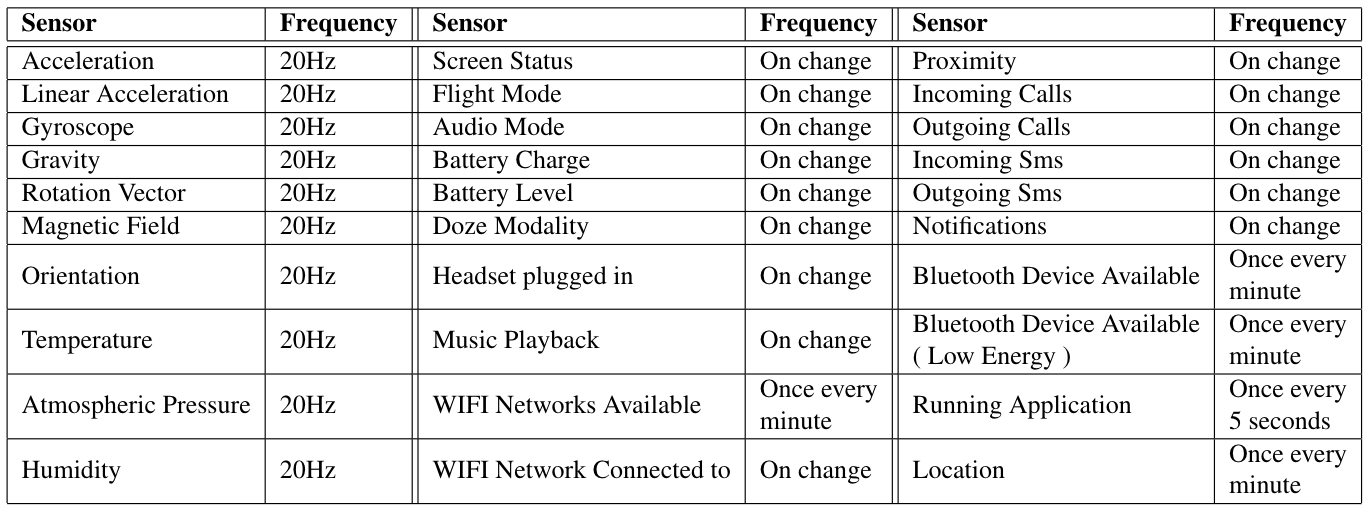}
\caption{Hardware and software sensor data collected together with their sampling rate. ``On change" means that the value of the sensor is collected only when it changes its value.}
\label{sensor data}
\end{figure}

\subsubsection{Q3: The chain of causal effects on the quality of the data}

To test the chain of events on the error, we use the position detected by the GPS of the smartphone when the user claims to be ``at home". As from Fig.\ref{Time diary} , ``at home" is one of the possible answers to the question "\textbf{Where are you?}", while the GPS is collected ever minute, as from the list of collected sensor data reported in
Fig. \ref{sensor data} (here the GPS is labeled as ``location").
We have selected the variable ``at home" for three reasons. The first is because we know the GPS position of the respondent's home
. The second is because home usually covers a very small area, differently from what is the case with other locations, for instance ``at the university”. The third is that that this is the type of question which would seem impossible to get wrong; it requires no thinking or reasoning of any kind, and it is also self-evident from the perceptual and habit point of view. Being at home is most likely the best known situation for everybody.

The variables used here are the same four used for (Q2) (see Section 3.1.2 above), that is, competence with iLog, reaction time, social context, mood and procrastination, plus the accuracy of the GPS (considered as an additional effect of the computing context). 
The resulting number of wrong answers turned out to be  substantial thus justifying the research hypothesis  motivating this research, that is, that it cannot be assumed that, modulo a minor number of local mistakes, all the user-provided answers are correct (see Section 2 on the Related Work). We had infact 730 answering errors out of a total number of 7624 questions, for an overall percentage of 9.6\% of wrong answers.

\subsection{Sample Recruitment and questionnaires scheduling}
\label{sec-33}

The data was collected as part of the Smart UNITN 2 project, as preliminarily approved by the Ethical Committee and GDPR Committee of the University of Trento, Italy. The project lasted for a total of four weeks (28 days) from the beginning of May to the beginning of June. The data collection was organized in five phases, as follows.

\vspace{0.1cm}\noindent
\textit{Phase 1: Process bootstrap.}
 A first short questionnaire was sent to a random sub-sample of 10006 students asking them if they regularly attended classes and if they had an Android smartphone. The sample was selected from the entire student population of University of Trento.

\vspace{0.1cm}\noindent
\textit{Phase 2: Sample set-up and profiling.}
For those who filled out the questionnaire and responded positively to both previous requests, the next step was an  invitation to participate in the survey.
The invitation explained the aim of the study and that students could choose to participate in the study for two or four weeks, and that
 they would receive a notification every half-hour in the first two weeks, and every two hours in the second two weeks. Students were also informed of the fact that various types of sensor data would be collected (see Fig.\ref{sensor data}).
As stated in literature \cite{S-2019-Keusch,aminikhanghahi2019context}, the willingness to participate in mobile data collection is strongly influenced by the incentive promised for study participation. A reward of 20 euros was promised to each participant for each of the two weeks of participation. In addition, each participant was informed that, at the end of the survey, there would be a lottery among those who responded to more than 75\% of the notifications; and that the lottery would assign three prizes of 100 euros for the first two weeks and three prizes of 150 euros for the second two weeks. The invitation also included a second questionnaire asking for additional information. The collected data was
 about the general characteristics of participants, their university experience, and the profiling of their procrastination syndrome. In this phase also the personal email plus the signed GDPR-compliant consent were collected.

\vspace{0.1cm}\noindent
\textit{Phase 3: Sample finalisation.}
The hiring process resulted into 1042 applicants. The response rate is perfectly aligned with other web surveys in which no reminders are sent. From these, those over the age of 25 were excluded in order to limit students with non-regular careers or who were close to the dissertation. A stratified sample of 318 students, proportional to the student population of each department of University of Trento, was drawn from the remaining 860 applicants. A second questionnaire was sent to the 318 sampled students, whose goal was to investigate their university life, their habits and their routines. All students who filled out the questionnaire and signed a second informed consent form were sent a password which allowed them to install iLog. 275 students completed the questionnaire and installed iLog.

\vspace{0.1cm}\noindent
\textit{Phase 4: Time diary data collection - first two weeks.} As declared in Phase 2, students received a notification, with the questions as from Fig.\ref{Time diary}, every half an hour. For each notification, participants had 720 minutes (12 hours) to provide an answer. After this period the question would be dropped and treated as a missing. Lastly, to reduce the burden, the user could stop data collection for 6 hours when going to sleep. 
Of the 275 students, only 237 answered the notifications at least once. In the first days 25 students dropped out, partly because of technical problems, partly because they were no longer interested in participating. At the end of the first two weeks, 184 students had provided valid responses to notifications for at least 13 out of 14 days.  

\vspace{0.1cm}\noindent
\textit{Phase 5: Time diary data collection - second two weeks.}.
Two days before the end of the first two weeks, a third questionnaire was sent out to ask about their iLog user experience. In this questionnaire, students were also asked whether they were willing to continue for the following two weeks. Of the 237 students, 202 declared their willingness to continue.  In this last two weeks, as declared in Phase 2, students received a notification every hour.
Of these students, 113 completed the task for at least 12 days and provided more than 100 valid answers.

\subsection{The Statistical Models}
\label{sec-32}

In the analysis of Q1 and Q2 we have used a multilevel discrete-time event history model (MLM) \cite{steele2008multilevel,S-2012-Tekle} and a Cox regression model \cite{allison2018event,steele2008multilevel,cox1972regression}. An MLM is a generalized linear model where repeated notifications over time (level 1) are nested within users (level 2) \cite{goldstein2011multilevel,singer2003applied} by modeling variability across upper-level units using random effects. Furthermore, MLM's allow to estimate the parameter attributes at multiple distinct levels. The most common example of MLM is that of a school, in which the competence of a student in a class is due in part of the student (first level, e.g., his/her ability) and in part is due to the class (second level, e.g., the attitude of the teachers of the class). 

For Q3, we have used both a multilevel structural equation path analysis model (MSEM) \cite{S-1986-Holland,hox2013multilevel,joreskog1979advances} and a standard SEM path analysis that integrates the evidence from the first two analyses into a single chain of causal events. In this case, reaction and completion time are influenced by the user features, while the response error is modeled as a random variable. For this reason, an MSEM has been chosen where reaction and completion time are formalized as a two-level model (responses nested to respondents) while the distance between the device and the user statement of being ``at home” is treated at the response level.

%% file: section/4.Reaction_time.tex
\begin{table}[htp!]
\centering
\begin{tabular}{lll}
\hline
\multirow{2}{*}{Features}            & Cox-regression      & Multilevel       \\ 
                                     & Coef B         & Coef B         \\ \cline{1-3}
\textit{(Spatial context): Where are you?}     &               &               \\
Home/Room                            & Reference     &    Reference           \\
Relatives Home                       & -0.061***     &   -0.100***            \\
House(friends/others)                & -0.021     &     -0.029          \\
University                           & 0.188***      &     0.277***          \\
Shop/Pub/Theatre                     & 0.028      &      0.013         \\
Work place                           & -0.234***     &      -0.331***         \\
Other place                          & -0.071***    &     -0.128***          \\
Outdoors                             & -0.218***     &    -0.324***          \vspace{0.2cm} \\
\textit{(Event context): What are you doing?}  &               &               \\
Personal care                        &  Reference             & Reference     \\
Eating                               &   -0.141***            & -0.193***     \\
Study alone or with others                                &  -0.174***             & -0.277***     \\
Class room lecture                               &   -0.130***            & -0.175***     \\
Social life/Break                 &  -0.137***             & -0.167***     \\
Watching YouTube TV, etc.            &   -0.080***            & -0.126***     \\
Social media, Phone call, chat            &  0.214***             & 0.311***      \\
Free time                            &   -0.298***            & -0.406***     \\
Work                                 &    -0.195***           & -0.271***     \\
Travel                               &   0.007            & 0.043    \vspace{0.2cm}    \\
\textit{(Social Context): Who are you with?}  &               &               \\
Alone                                & Reference     &       Reference        \\
Friend(s)                            & -0.188***     &    -0.292***           \\
Relative(s)                          & -0.046***     &     -0.079***          \\
Classmate(s)                         & -0.222***     &    -0.329***           \\
Roommate(s)                          & 0.025       &   0.018            \\
Colleague(s)                         & -0.238***     &     -0.396***          \\
Partner                              & -0.291***     &   -0.432***            \\
Other                                & -0.575***     &  -0.834***        \vspace{0.2cm}     \\
\textit{(Time Context): When questions sent}    &               &               \\
Sunday                                & Reference     &       Reference        \\
Monday                               & 0.089***      & 0.122***      \\
Tuesday                              & 0.115***      & 0.154***     \\
Wednesday                            & 0.068***      & 0.076***      \\
Thursday                             & 0.131***      & 0.172***      \\
Friday                               & 0.032**       & 0.049**      \\
Saturday                             & -0.022       & -0.032       \\
Study day                     & -0.076***     & -0.123***     \\
Study day$^2$                    & 0.003***    & 0.005***      \\
Question delivery delay time (sec.)        & 0.0002***     & 0.0003***   \vspace{0.2cm}  \\
\textit{User Characteristics}             &               &               \\
Mood                                 & 0.060***      & 0.092***      \\
Procrastination                      & -0.010**      & -0.021**     \\
\textit{var(cons[user])}                           &      & 0.643***     \\ 
\textit{var(cons[user\textgreater notid])}                              &      & 0.604***     \\ 
\textit{Observations}                              & 58340     & 2,565,159     \\ 
\textit{Number of groups}                              & 158     & 158     \\ 
\bottomrule
\end{tabular}
\vspace{0.2cm}
\caption{Cox Regression model and Multilevel discrete time model with random intercept on the reaction time. (Note: (*) p\textless 0.1; (**) p\textless 0.05; (***) p\textless 0.01).}
\label{H1.1}
\end{table} 

\subsection{Q1: Reaction time}
\label{H1}

The results are reported in Tab.\ref{H1.1}.
The median survival time of the reaction time is 20 minutes, i.e., fifty per cent of all notifications receive a response in 20 minutes or less. As assumed (see Section \ref{sec-rq}), the reaction time is influenced by both the historical context and the characteristics of the user. For example, the median reaction time of responses varies from a minimum of 10 minutes when the user is involved in social media/phone/chat activities to a maximum of 27 minutes when involved in free time activities (Log-rank test. $\chi^2(9) = 1046.69$, p\textless 0.05). In a social context, the median reaction time varies from 16 minutes when the user is alone to 33 minutes when in a social setting (Log-rank test. $\chi^2(7) = 635.38$, p\textless 0.05). With respect to location, the median reaction time ranges from 15 minutes when the user is at university to 32 minutes when outdoors. (Log-rank test. $\chi^2(8) =  708.50$, p\textless 0.05). 

To test the net effects of all the features influencing the reaction time considered here, we have run two regression models (Tab. \ref{H1.1}), a Cox-regression model with user notification shared probability and a Multilevel discrete time model. For both models we have performed a likelihood ratio test with the null model \cite{bolker2009generalized}. The results show that the two models are both statistically significant ($\chi^2(34) =  3437.34$, p\textless 0.05; $\chi^2(47) =  5819.65$, p\textless 0.05) and capture the same meaning (i.e., the probability to react at a notification at time t). The parameters differ slightly, but the sign and interpretation meanings are exactly the same. Moreover, with the Cox regression we can also estimate that the model explains 5.0\% of the variance in accuracy ($R_D^2= 0.0462$;) \cite{royston2006explained}. This means that by using the independent variables listed in Table \ref{H1.1}, we can explain 5.0\% of fluctuation in reaction time.

%% file: section/5.Completion_time.tex
\subsection{Q2: Completion time}
\label{H2}

The results are reported in Tab.\ref{H2a}.
The median survival time of the completion time is 9 seconds, i.e., fifty per cent of all notifications are filled in 9 seconds or less. As assumed, also the completion time is influenced by both the historical context and the characteristics of the user. For example, the median completion time varies from a minimum of 7 seconds when the user is involved in study (study alone or with others) or lesson (classroom lecture) activities to a maximum of 11 seconds when involved in free time activities (Log-rank test. $\chi^2(9) = 5478.37$, p\textless 0.05). In a social context, the median time varies from 8 seconds when the user is alone to 10 seconds when with the partner (Log-rank test. $\chi^2(7) = 1682.50$, p\textless 0.05). With regard to location, the median time ranges from 7 seconds when the user is at university to 12 seconds when Shop/Pub/Theatre. (Log-rank test. $\chi^2(8) = 2072.78$, p\textless 0.05). 

\noindent
To disentangle the effects of different contexts on completion time, we also run a Cox regression model and a multilevel discrete-time model here (Tab.\ref{H2a}). The likelihood ratio test with the null model provide evidence that the models are both statistically significant ($\chi^2(8) = 3683.55$, p\textless 0.05; $\chi^2(19) = 8457.39$, p\textless 0.05) and show the same meaning. As for Q1, the parameters differ only slightly, but the sign is exactly the same. Moreover, with the Cox regression, we can also estimate that the model explains 4.1\% of fluctuation in completion time ($R_D^2= 0.0407$;).

\begin{table}[htp!]
\centering
\vspace{-0.1cm}
\begin{tabular}{lll}
\toprule
\multirow{2}{*}{Features}            & Cox-regression      & Multilevel       \\ 
                                     & Coef B         & Coef B         \\\cline{1-3} 
\textit{Event context:} Activity    &   0.038***                   & 0.104***      \\
\textit{Social context:} Alone vs not alone      & 0.267***                    & 0.637***   \\
\textit{User Characteristics:} Mood  \ \ \ \           & 0.039***                  & 0.095***    \\
\textit{User Characteristics:} Procrastination\ \ \ \ \  & -0.006*    & -0.015*   \\

Study time         & 0.048***             & 0.131***    \\
Study time$^2$         & -0.002***            & -0.005***   \\
Reaction time (min.)       & -0.0002***         & -0.001***   \\
Pending notification (count)      & 0.044***           & 0.093***    \\
Theta/constant         & 0.0731***                                  & -14.704***  \\
\textit{Observations}         & 58,340                                  & 2,565,159  \\
\textit{Number of groups}          & 158                                  & 158 \\
 \bottomrule
\end{tabular}
\vspace{0.2cm}
\caption{Cox regression model and Multilevel model with random intercept on the completion time. (Note: (*) p\textless 0.1; (**) p\textless 0.05; (***) p\textless 0.01).}
\label{H2a}
\end{table}

%% file: section/6.Answer_correctness.tex
\begin{figure}[htbp!]
\centering 
\subfigure[Reaction Time.]{
\label{Reaction Time}
\includegraphics[width=0.49\textwidth]{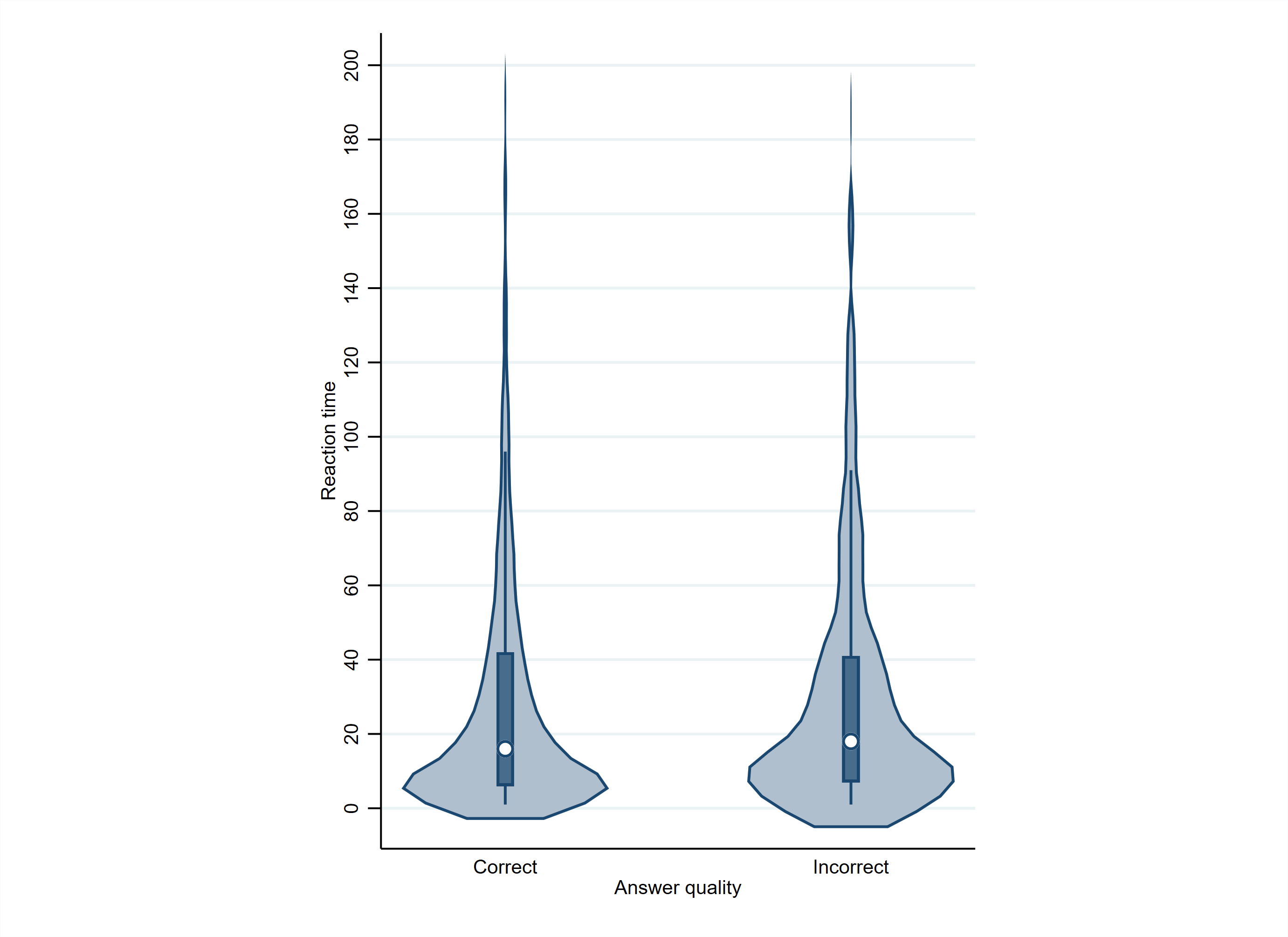}}
\subfigure[Completion time.]{
\label{Completion time}
\includegraphics[width=0.49\textwidth]{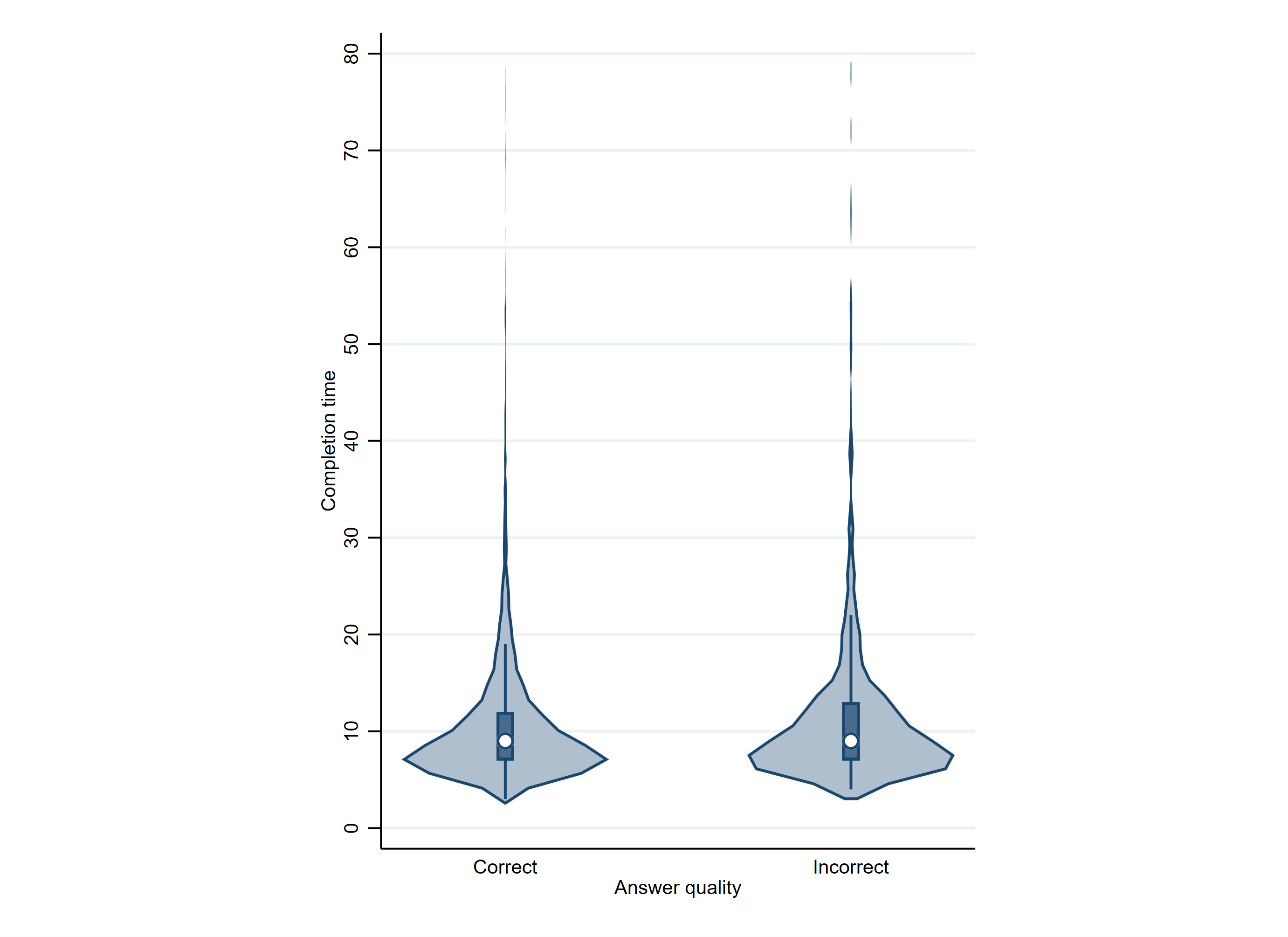}}
\caption{Which impact on the answer quality?}
\label{impact}
\end{figure}

\subsection{Q3: Answer correctness}
\label{sec:H3}

Fig. \ref{Reaction Time} reports a first exploratory analysis of the effects of reaction and completion time on the answer quality. This analysis shows that there is a statistically significant negative impact on the response quality from both the reaction time (Fig. \ref{Reaction Time}) and the completion time (Fig. \ref{Completion time}). While the first result was expected, the second shows that the negative effects of time on the completion time (e.g., memory errors) are bigger that the positive effects (e.g., increased time in the computation of the answer), see the discussion in Section \ref{sec-rq}. In fact, a comparison of the two groups of correct and incorrect answers, where a distance greater than 50 meters is considered an incorrect answer, provides the following values:

\begin{itemize}
\item \textit{Reaction time}
\begin{itemize}
\item Mean : Correct (38 minutes), Incorrect (43 minutes) (Fisher $F= 5.02$, p\textless 0.05); \item Median survival reaction time: Correct (17 minutes), Incorrect (19 minutes) (Log rank test: $\chi^2(1) = 4.93$, p\textless 0.05; Non-parametric equality-of-medians test: $\chi^2(1) = 3.73$, p\textless 0.05) \cite{mann1947test}.
\end{itemize}  
\item \textit{Completion time}
\begin{itemize}
\item	Mean : Correct (11.0 seconds), Incorrect (11.8 seconds) (Fisher $F= 4.73$, p\textless 0.05); 
\item Median survival completion time: Correct (9 seconds), Incorrect (9 seconds) (Log-rank test: $\chi^2(1) = 4.36$, p\textless 0.05; Non-parametric equality-of-medians test: $\chi^2(1) = 5.17$, p\textless 0.05).
\end{itemize}  
\end{itemize} 
\noindent
For both reaction and completion time the mean time for correct answers is lower than that of incorrect answers. This applies also to the median survival reaction time, while the median survival completion time is the same for correct and incorrect answers.

\begin{figure}[htp]
\centering
\includegraphics[scale=0.80]{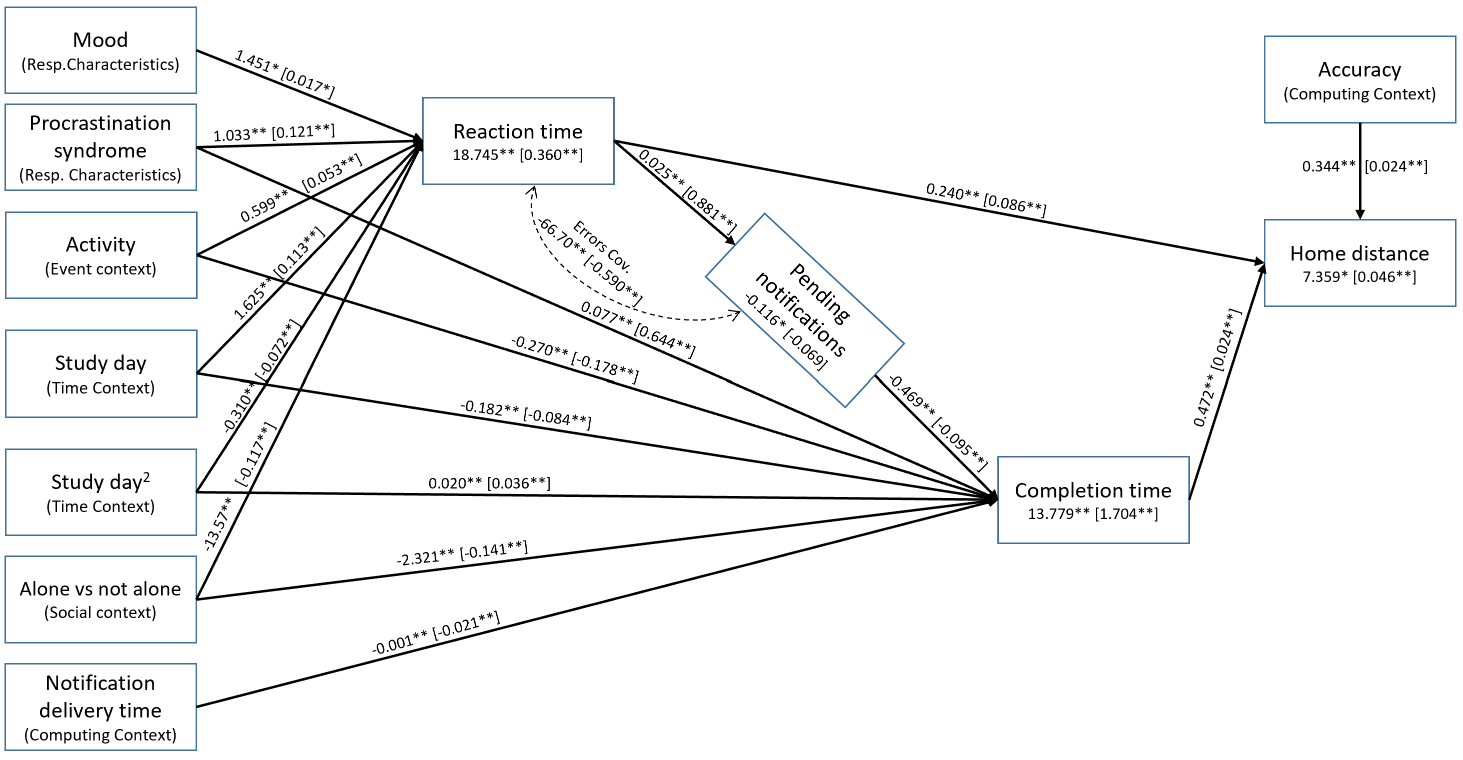}
\caption{Chain of errors: Multilevel structural equation model and a Structural Equation model.(Note: Between square brackets are reported the standardised parameters (i.e., ``Coef B") of the SEM model.)}
\label{Chain of errors}
\end{figure}
\noindent
However, reaction and completion are not independent of one another, but are links of the same chain. 
As from Fig. \ref{Chain of errors}, the two models we have developed, i.e., a Multilevel Structural Equation Model (MSEM) \cite{radu2018multimodal} and a Structural Equation Model or Path Analysis (SEM) (see Section \ref{sec-rq}), both support the theoretical model depicted in Fig. \ref{ The theoretical model} and the idea of a chain of events on answer quality \cite{S-2019-Read}\footnote{Due to the correlation between the reaction time and the number of pending notifications, the covariance between the errors of these two variables has been defined in the two models.}. In fact, the input variables are (mostly strongly) relevant and the fit indices show that the model fits the observed data very well (RMSEA= 0.022; SRMR=0.013; chi2=95.41 (20); $R^2=0.127$). The model shows a 13.0\% of fluctuation in the answer quality. Unfortunately, multilevel models present challenges in constructing fit indices because there are multiple levels of hierarchy to account for in establishing goodness of fit. As a result, there is a lack of consensus on suitable fit indices for multilevel models \cite{comulada2021calculating}.

As shown in Fig. \ref{Chain of errors}, context history, burden, technology, and personality aspects play a different role in ``when” the user decides to respond and in ``how” the user fills out the notification. In a chain of causal effects, reaction time (when) and completion time (how) have a direct effect on the quality of the data collected and the associated errors. In the causal model in Fig. \ref{Chain of errors}, the ``when” influences both the question-answer process and the quality of the data through the number of notifications to be filled in when the user begins to respond. In turn, this is influenced by the burden, repetitiveness of the activity, and compliance with the research protocol, which varies, according to the subject's personality characteristics, such as procrastination and daily mood, and the social context. The ``how,” is influenced by the burden, distraction due to the presence of other people, the repetitiveness of the activity, and the subject's personality level of procrastination, and directly influences the response error. In a nutshell, burden, activities, personality idiosyncrasy and cognitive aspects have a direct effect on the user behaviour and in turn the user behaviour has a direct effect on the error.

Finally, the overall conclusion is that \textit{reaction time is the key variable with the highest impact on the answer quality}. First of all, as from Fig. \ref{impact}, the reaction time is much higher than the completion time, where the mean of the former is 38 minutes for correct answers and 43 minutes for wrong answers, while that of the latter is 11 seconds for correct answers and 11.8 seconds for incorrect answers. Second, reaction times of the order of half-hours or longer largely facilitate memory errors and, therefore, wrong answers. It is also worthwhile noticing the relatively small difference between the median completion time for correct and incorrect answers (from 11 to 11.8 seconds).

%% file: section/7.Description.tex
\section{Take-away lessons}
\label{sec:des}

The goal of this section is to draw some conclusive remarks about the general take-away lessons. Section \ref{sec-fr1} analyses the role of the key exogenous factors (i.e., situational and temporal context) as well of the endogenous factors (i.e., the general user characteristics) on the answer quality. Section \ref{sec-fr2} analyses the chain affect from reaction time to completion time. Finally, Section \ref{sec-fr3} provides a set of general recommendation, based on the results from this project but also on results from the literature, about how to execute future ESM data collections and experiments.

\subsection{The role of exogenous and endogenous factors on the answer quality}
\label{sec-fr1}

Although we trust the quality of our memories, and so has done the previous work on EMA/ESM compliance (see Section \ref{H1}), research on autobiographical memory teaches us that memory can be unreliable \cite{bradburn1987answering,S-2000-Tourangeau,stone2007science}. Our recollections are not just inaccurate, they are often systematically biased \cite{shiffman2008ecological}. The more time elapses from what we want to recall, the greater the risk of making mistakes.
Everything pivots around the effects that the different contexts have on time and, in turn, on memory and the cognitive process concerning the response quality. Based on the state of the art (Section \ref{sec:related}) and on the analysis we have performed (Section \ref{sec:new}), we present below a detailed view of how all these aspects have a direct and/or indirect significant effect on the entire response process and, in turn, how these, taken together, influence the quality of responses.

\subsubsection{The situational context}

\begin{figure} 
\begin{minipage}[htp]{0.60\linewidth} 
\includegraphics[width=2.6in]{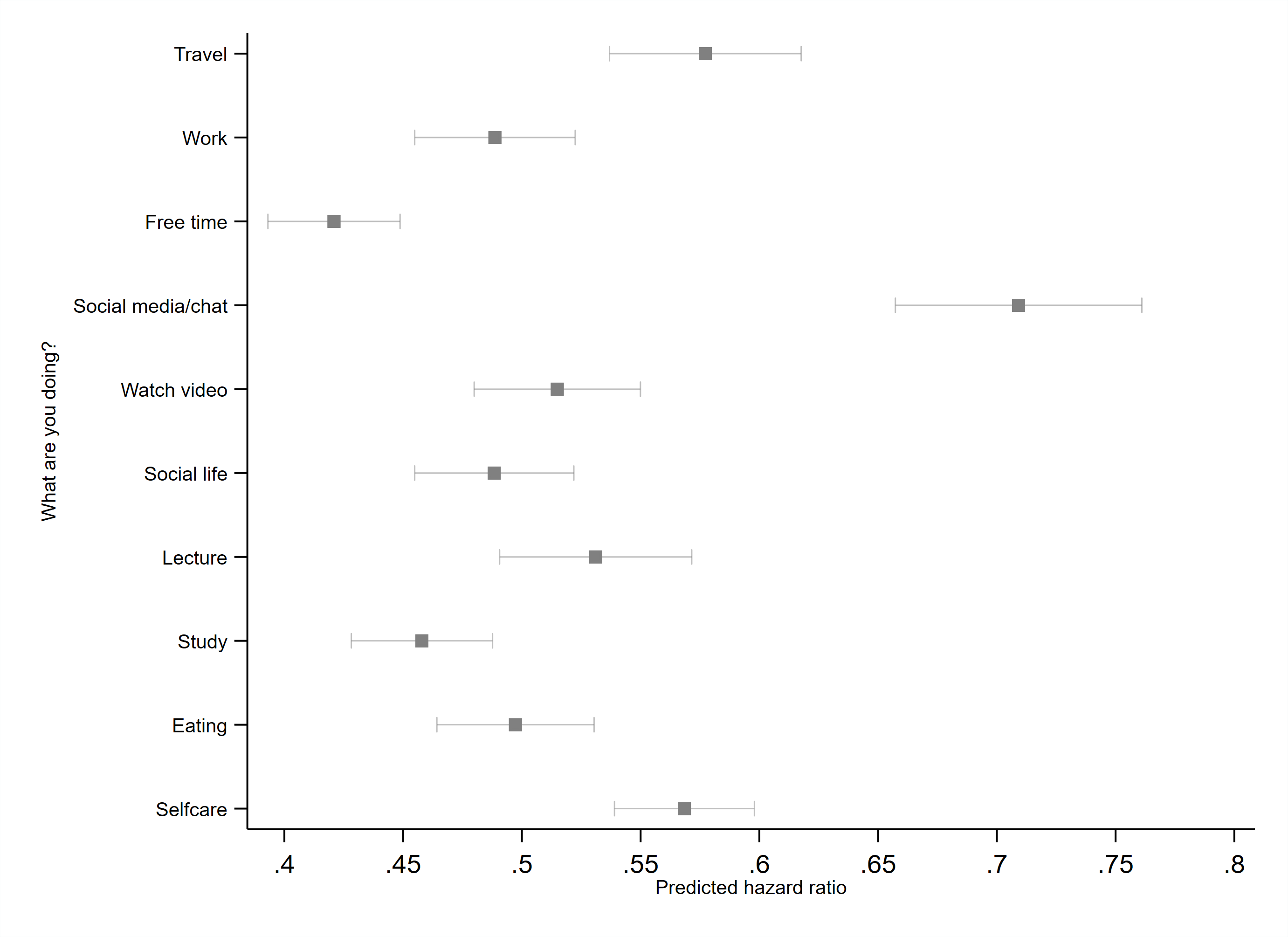} 
\caption{Activity effect on reaction time} 
\label{activity} 
\end{minipage}%
\begin{minipage}[htp]{0.60\linewidth} 
\includegraphics[width=2.6in]{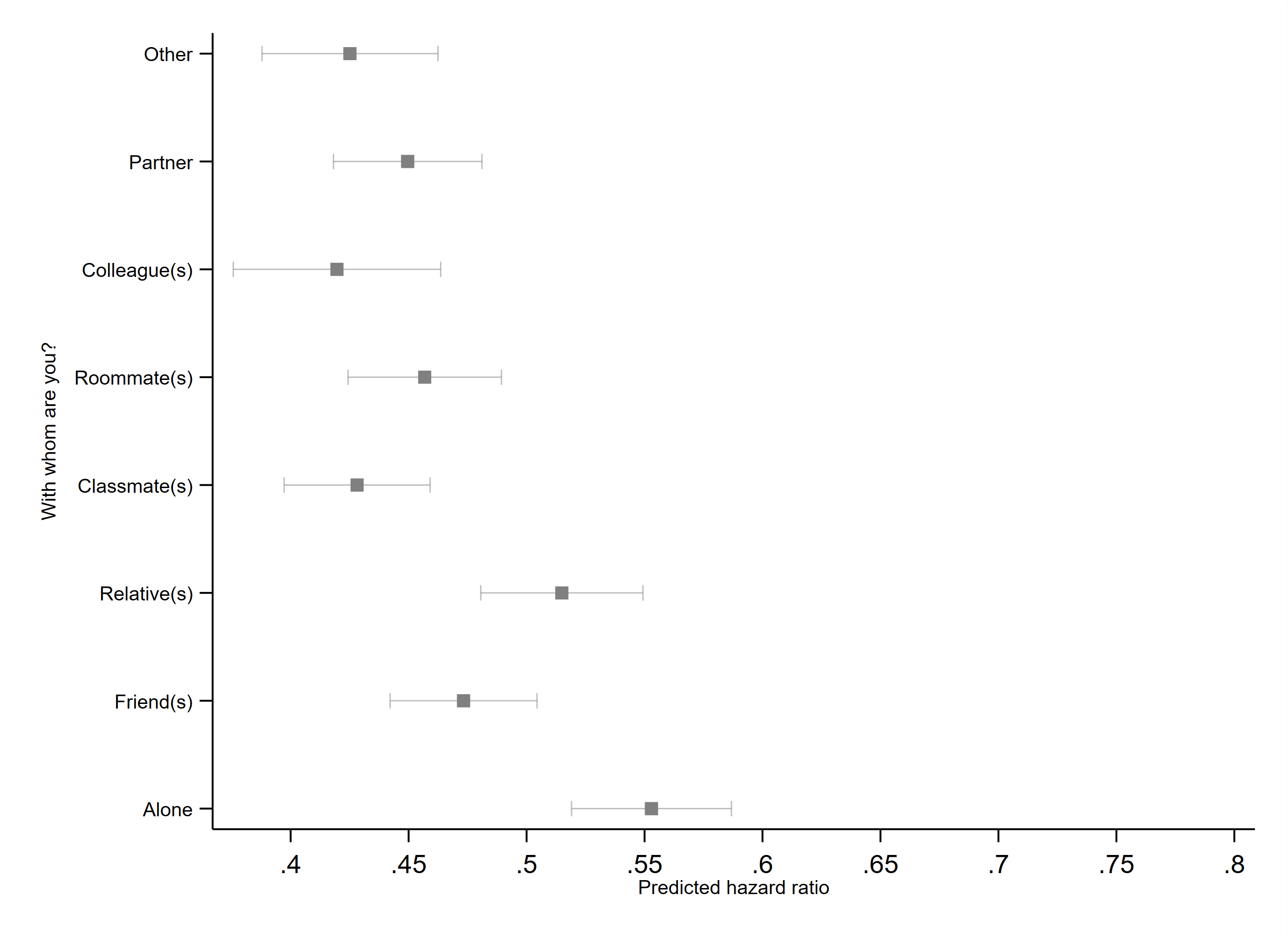} 
\caption{Social relation effect on reaction time} 
\label{Social relation} 
\end{minipage} 
\end{figure}

\begin{figure} 
\begin{minipage}[htp]{0.60\linewidth} 
\includegraphics[width=2.6in]{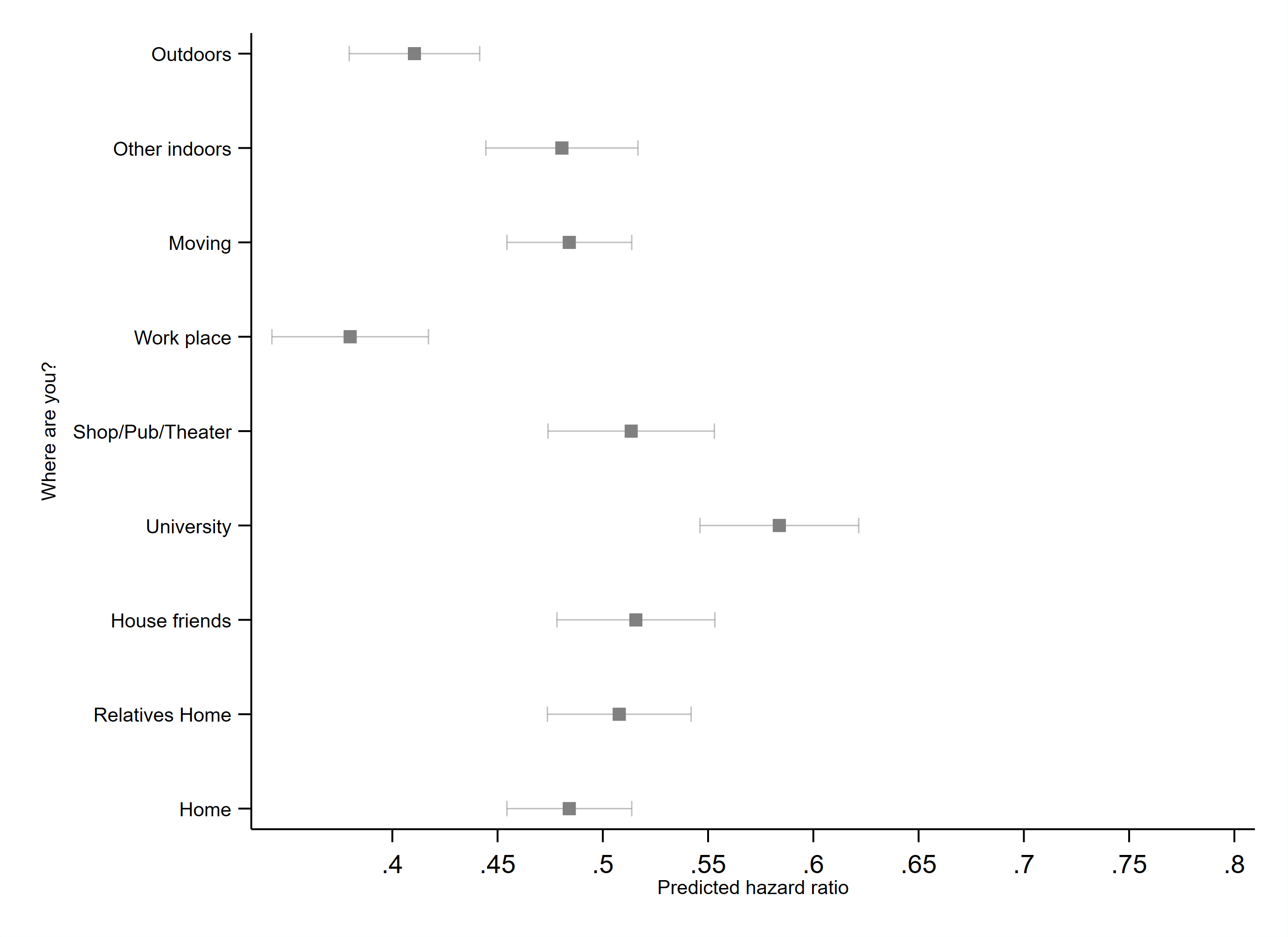} 
\caption{Location effect on reaction time} 
\label{Location} 
\end{minipage}%
\begin{minipage}[htp]{0.60\linewidth}  
\includegraphics[width=2.6in]{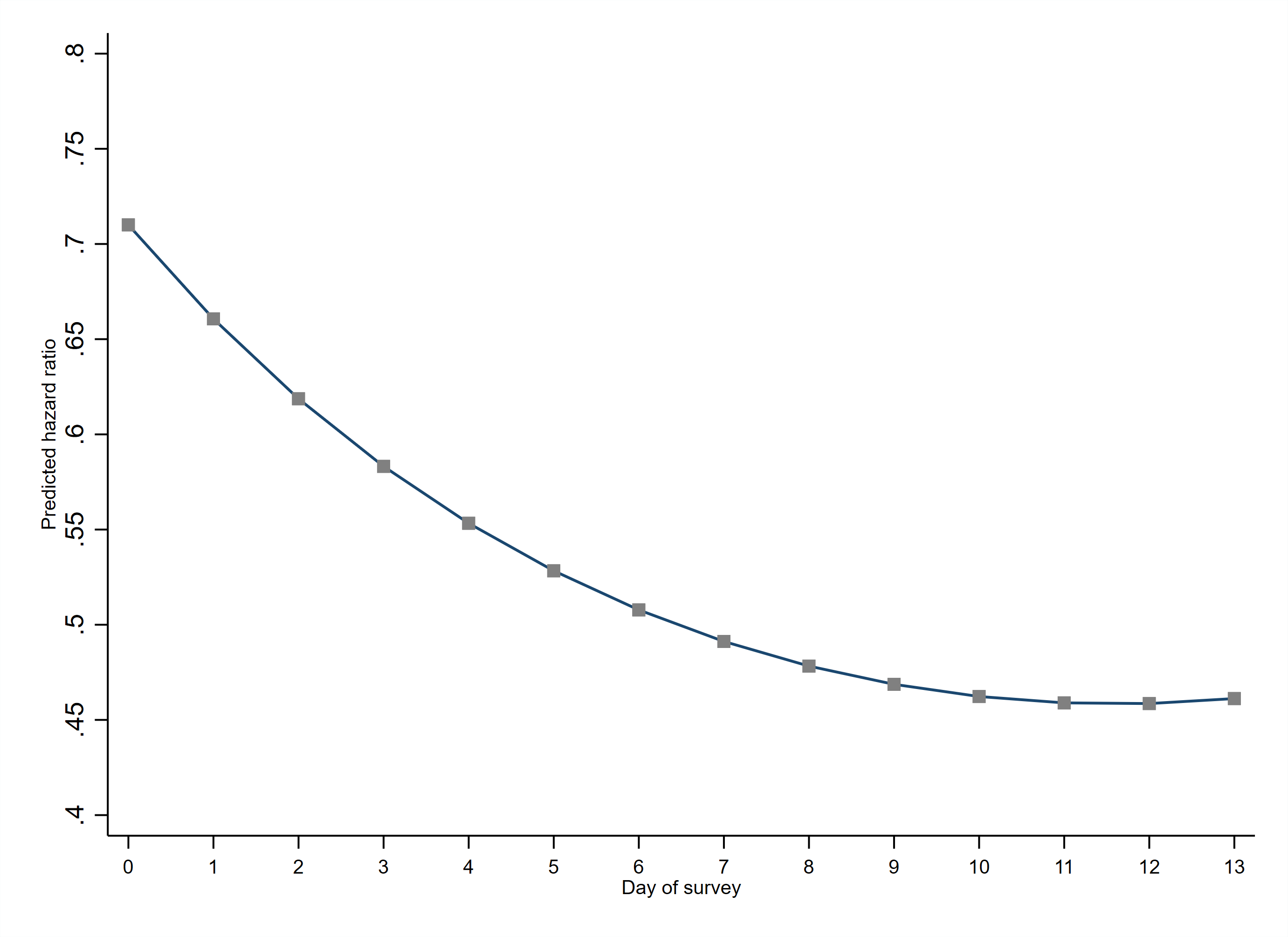} 
\caption{Survey Day effect on reaction time}
\label{Survey Day} 
\end{minipage} 
\end{figure}

Location, activity, and social context (Tab. \ref{H1.1}) influence in a very different way the reaction time (Fig. \ref{activity}, \ref{Social relation}, \ref{Location}). Thus, e.g., being ``at university", ``alone", connected with ``social media" significantly increases the likelihood that the user will respond very early to the incoming notification. On the other hand, being ``outdoors", in ``leisure", with a ``partner" can significantly increase the delay with which the user decides to respond (Fig. \ref{user profile}). As it emerges from this model, one cannot simply detect whether the phone is \textit{on/off} \cite{wang2018tracking}, or whether the subject is changing activity \cite{H-2017-Pielot,S-1986-Holland}. The interaction between location, activity and social relation, combined with time, plays a decisive role on the reaction time. 

\begin{figure} 
\begin{minipage}[htp]{0.60\linewidth} 
\includegraphics[width=2.6in]{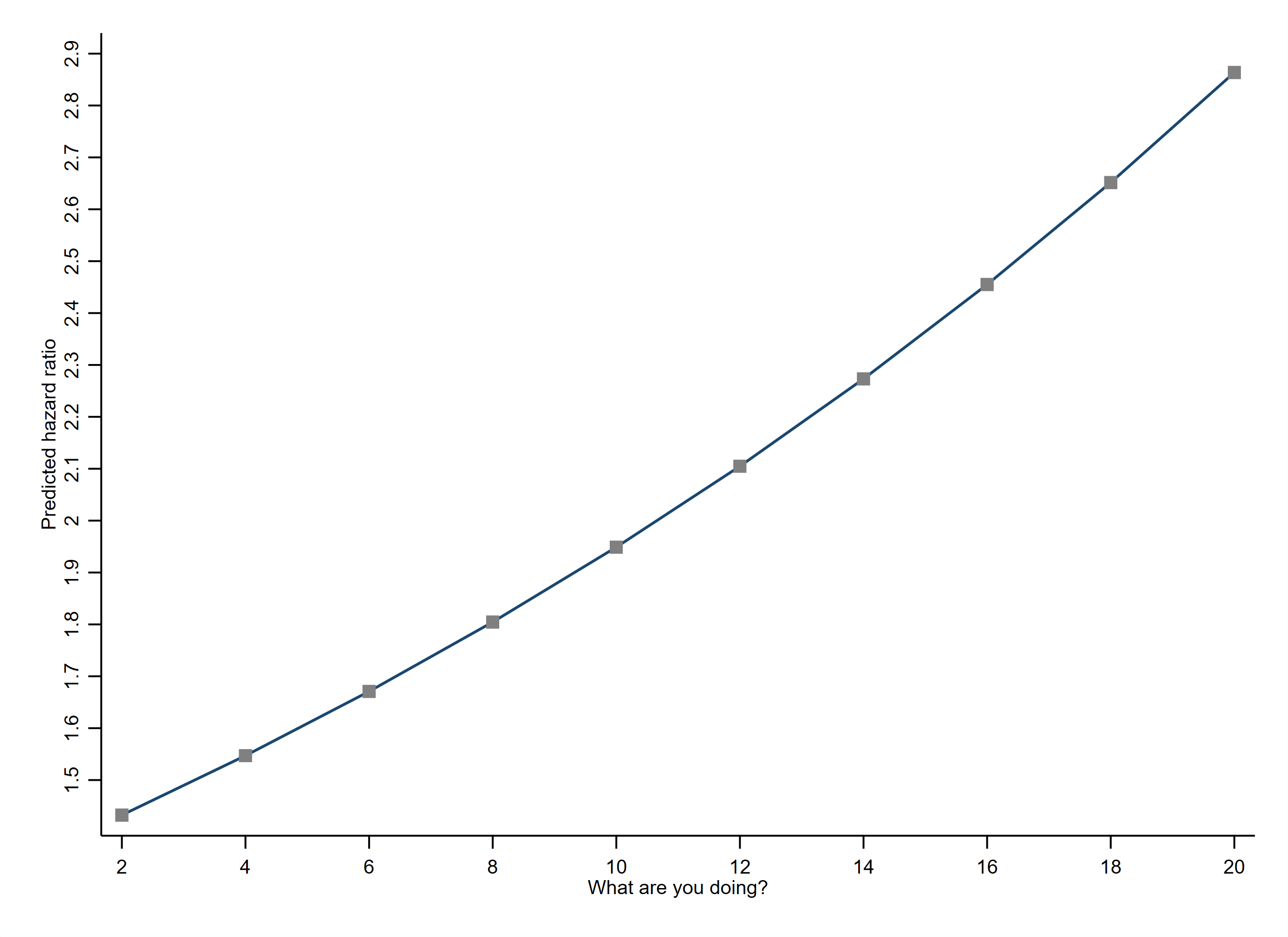} 
\caption{Activities probability effects on completion time} 
\label{Activities ct} 
\end{minipage}%
\begin{minipage}[htp]{0.60\linewidth} 
 
\includegraphics[width=2.6in]{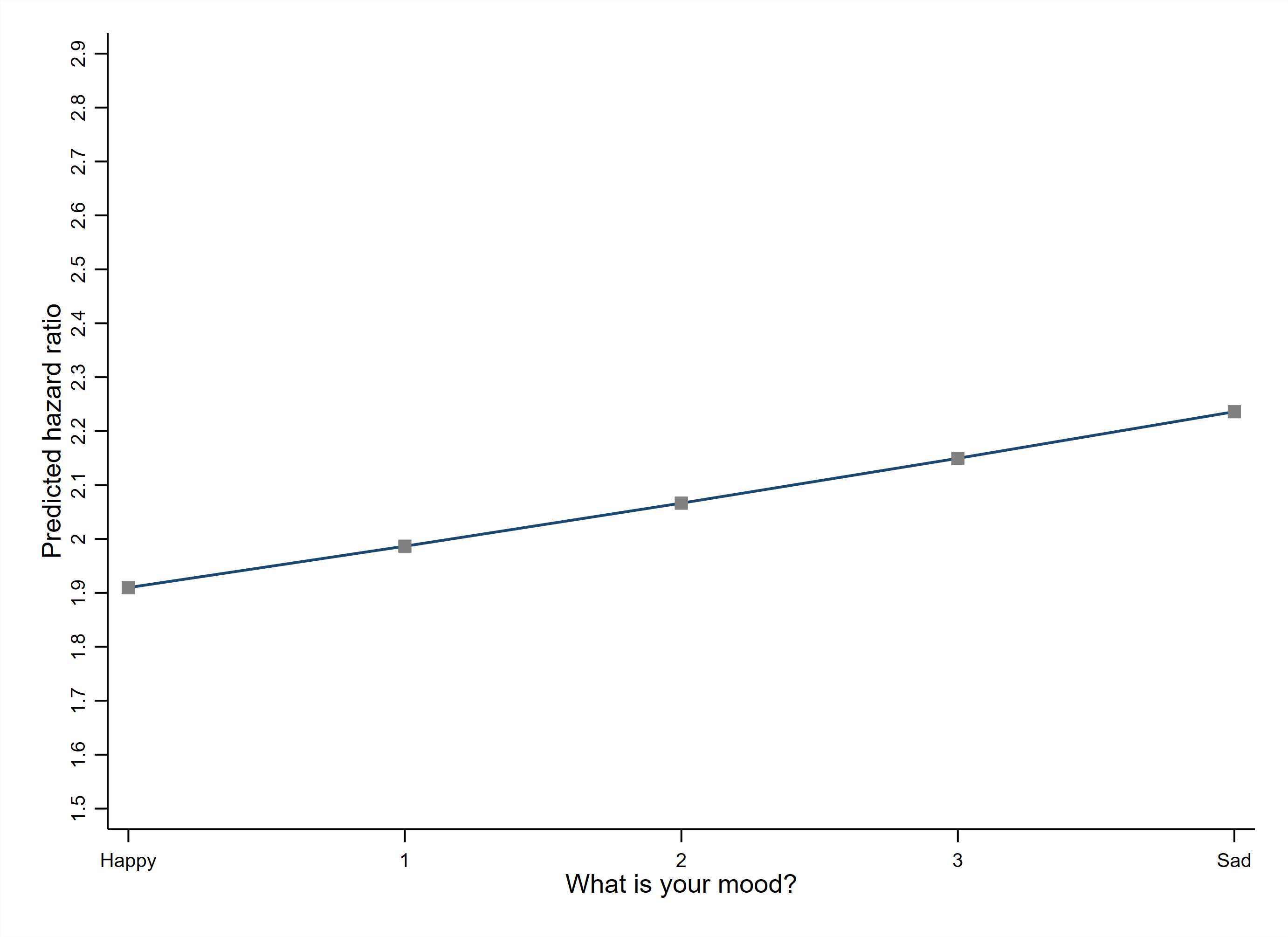} 
\caption{Mood effects on completion time} 
\label{Mood ct} 
\end{minipage} 
\end{figure}

Activities and social context also influence completion time at two different levels (Tab. \ref{H2a}). The first level, as expected, is that the repetition of  events (fig. \ref{Activities ct}) has the effect of reducing the compilation time. In this case, the cause is the cognitive training process in which, notification after notification, the user learns how to classify different activities in a limited list of alternatives. The greater the probability that an event occurs, the greater the probability that the user remembers its position. The second level is that externally induced distraction effect affects time; thus being alone decreases completion time (Tab. \ref{H2a}). That is, when  alone the user finds it easier to concentrate on the answer.

\begin{figure} 
\begin{minipage}[htp]{0.60\linewidth}  
\includegraphics[width=2.6in]{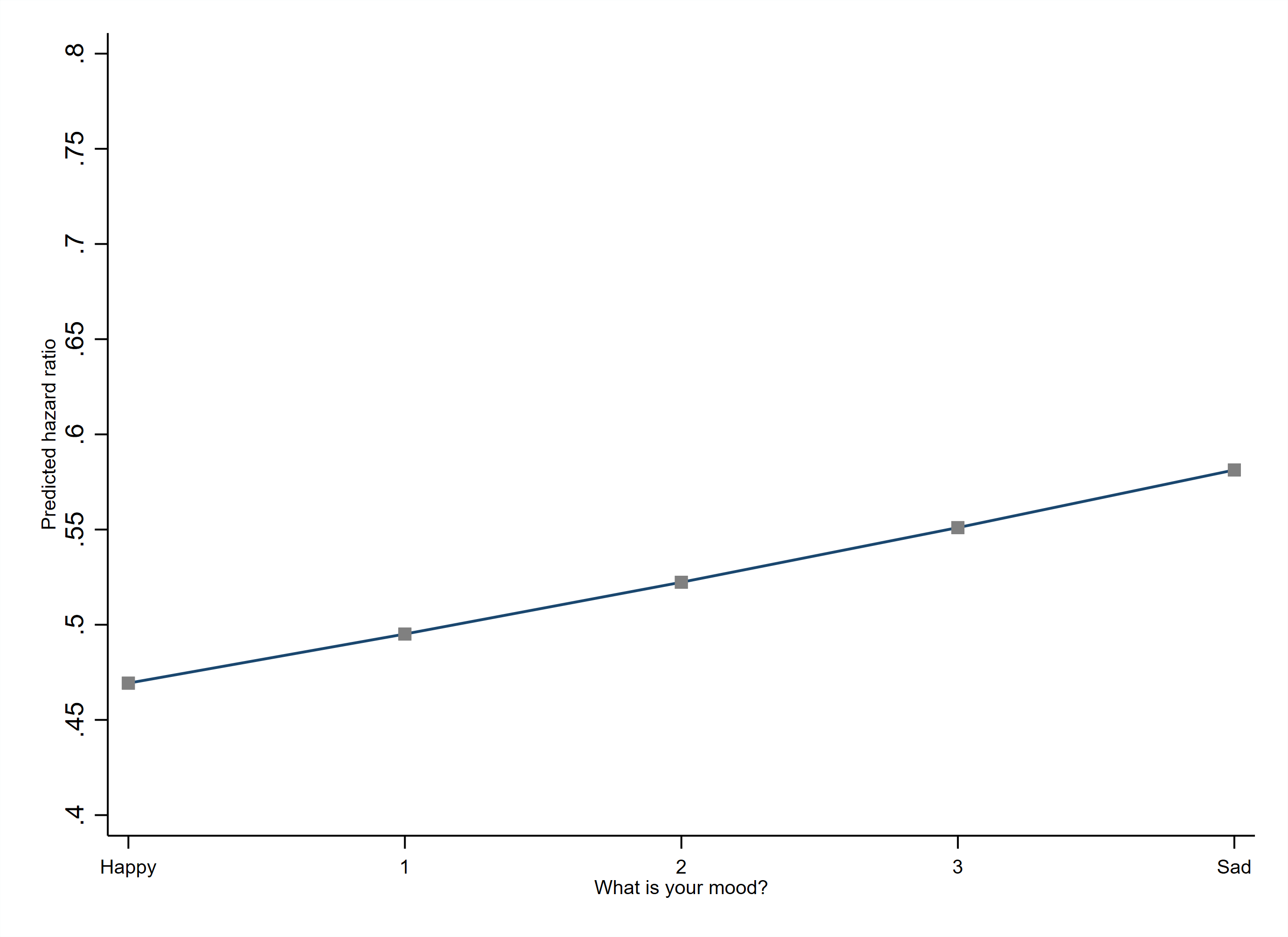} 
\caption{Mood effect on reaction time} 
\label{Mood} 
\end{minipage}%
\begin{minipage}[htp]{0.60\linewidth}  
\includegraphics[width=2.6in]{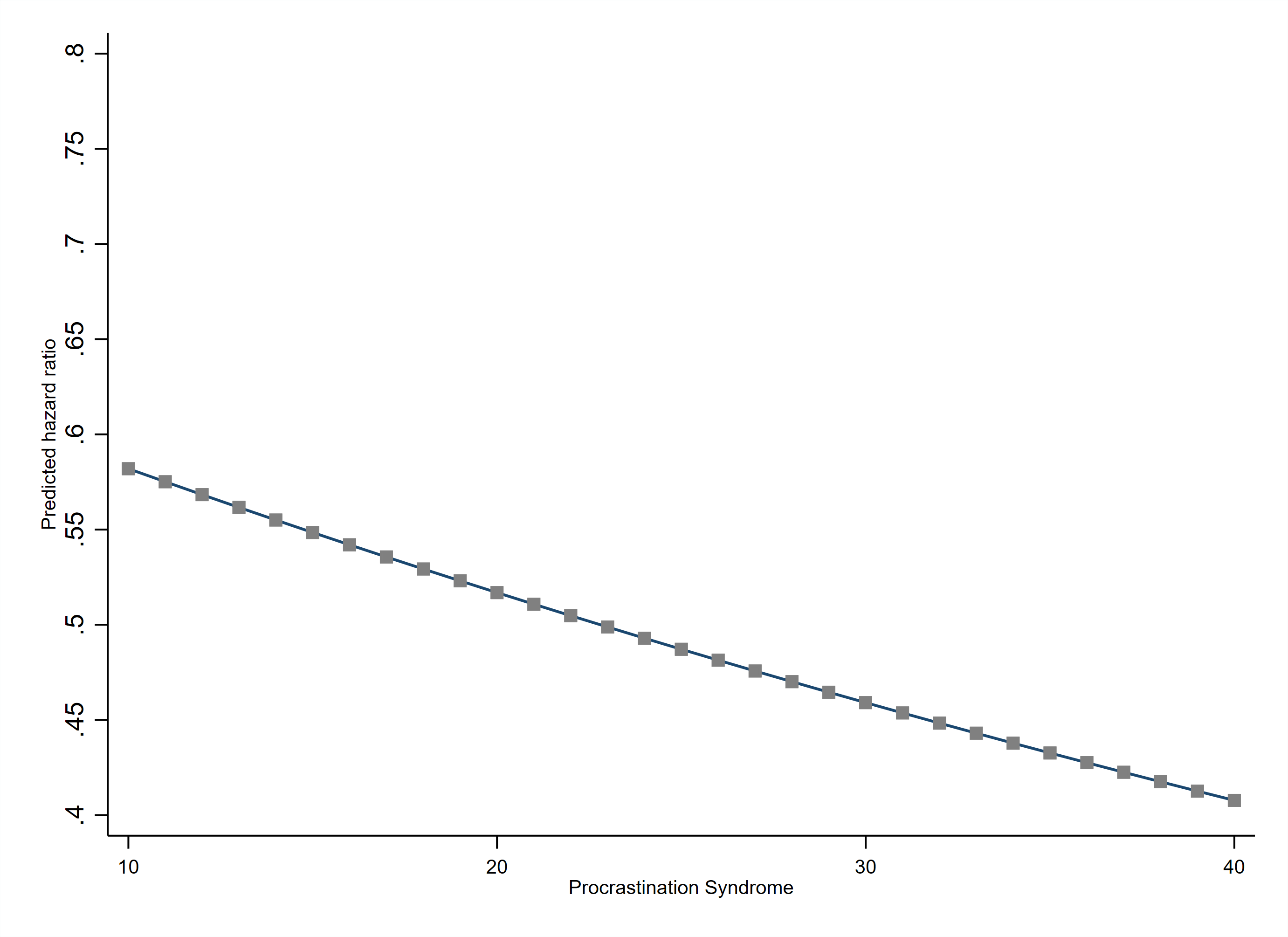} 
\caption{Procrastination effect on reaction time} 
\label{Procrastination} 
\end{minipage} 
\end{figure}

\subsubsection{The temporal context}

The time context matters. Over time (Fig. \ref{Survey Day}), the user tends to increase the reaction time. However, this delay increases rapidly in the first few days and then seems to find a balance (trade-off) between the task and the time at which it is performed. 

\begin{figure} 
\begin{minipage}[htp]{0.60\linewidth} 
\includegraphics[width=2.6in]{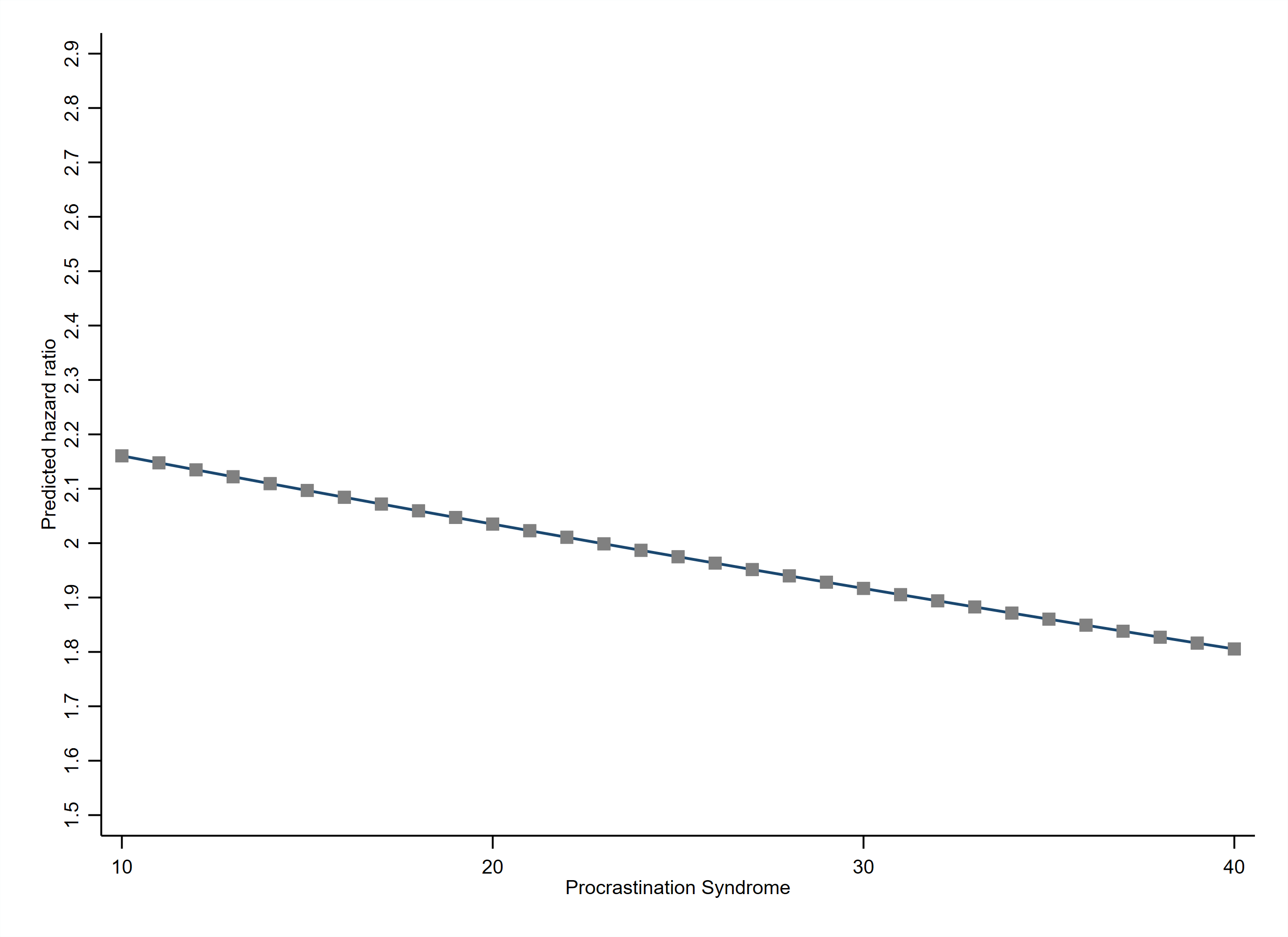} 
\caption{Procrastination effects on completion time} 
\label{Procrastination ct} 
\end{minipage}%
\begin{minipage}[htp]{0.60\linewidth} 

\includegraphics[width=2.6in]{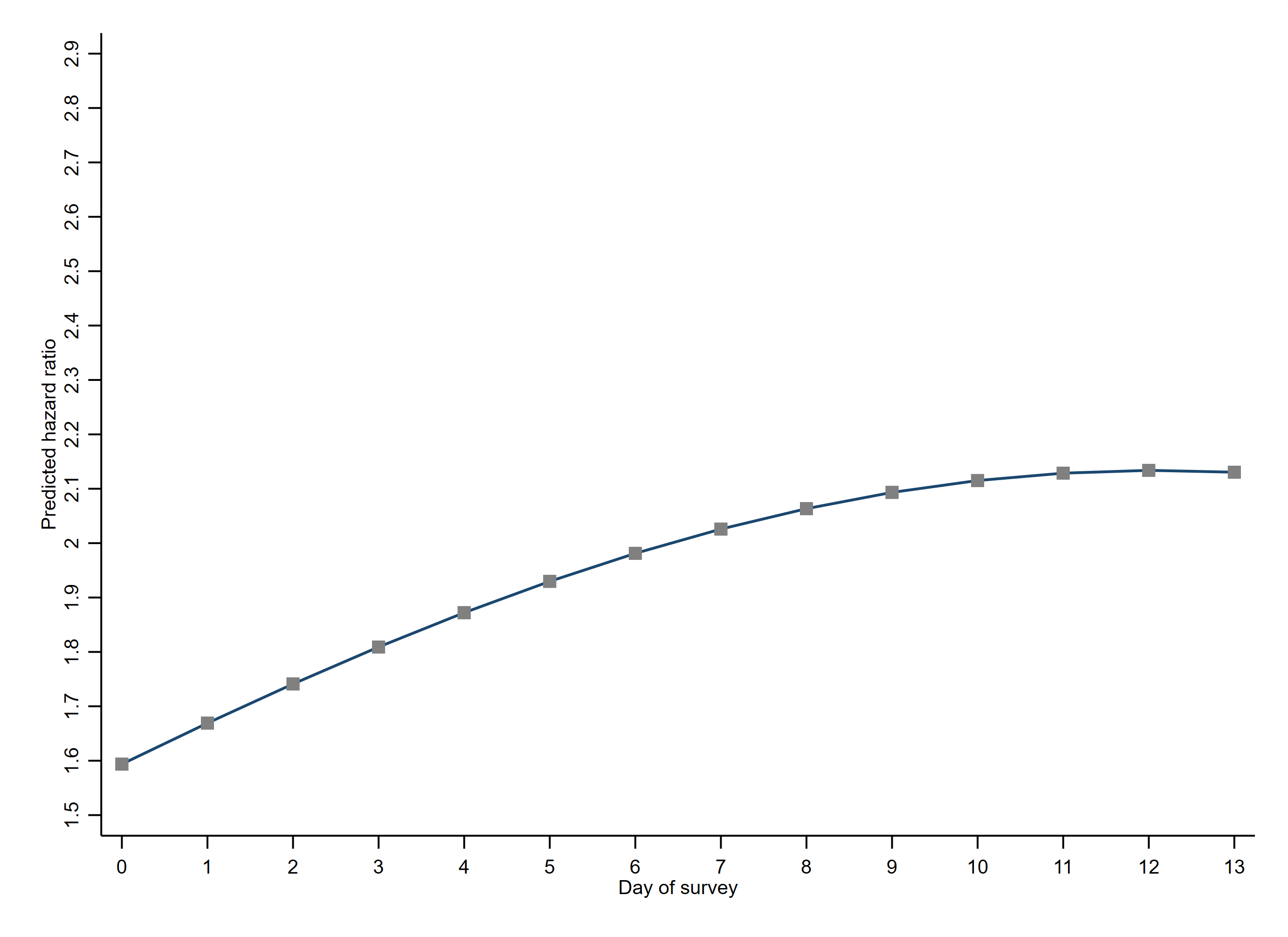} 
\caption{Survey day effects on completion time} 
\label{Survey day ct} 
\end{minipage} 
\end{figure}

Opposite is the effect of app usage on the completion time. In this case, the user tends to reduce the completion time over time, see Fig. \ref{Survey day ct}. 
However, we can assume that this effect of the learning process is only related to the first observation period. In fact, the indirect effect on the quality of responses seems to become constant after one week (Fig. \ref{Chain of errors}). In other words, over time the user finds a balance between the frequency of the notifications sent by the server (every half an hour), the burden of the task, the compliance with the research, his life routine and his cognitive ability. This means that if we want to observe the user over a long period, we must replace the day of the survey with a more sensitive routine of response. 

Finally, there is a third type time, that we can call the social time \cite{paetzold2008review}. This type of time refers to the daily activities of the students and to how they change during the week. Students do not necessarily attend classes every day nor do they follow precisely the same routine. Looking at the days of the week, the response delay behaviour is opposite to what one might expect. Longer reaction times do not occur during the period of maximum academic activity, i.e., weekdays, but rather during the weekend, i.e., Saturdays and Sundays. Taking time for oneself and relaxation seems to reduce the attention to the task. 

\subsubsection{User characteristics}
Sex and age were found to play no statistically significant or even moderated effects.
Procrastination increases the delay for both reaction time and completion time (Fig. \ref{Procrastination} \& \ref{Procrastination ct}) while a sad mood reduces it (Fig. \ref{Mood} \& \ref{Mood ct}). This confirms the literature \cite{S-2013-Forgas} which states that a lower value of mood corresponds to an increase in attention and also memory. Notice that, from the Q3 model in Section \ref{sec:H3}  (Fig. \ref{Chain of errors}), the mood has a significant direct effect only on reaction time and not on completion time. 
However, Steel \cite{S-2007-Steel} argues that there is a relationship between mood and procrastination.   Procrastination is a way of temporarily evading anxiety and therefore to improve mood in the shorter term, but with a negative effect in the longer term (because of the increase of tasks not completed). This opens the possibility of a depressive spiral where depression may lead to procrastination and then to a bad mood. Therefore, in an interactive survey process with the smartphone, there is the possibility that mood influences memory, accuracy, motivation, and therefore the completion time.

\subsection{The chain effect of reaction and completion time on the answer quality}
\label{sec-fr2}

Completion and reaction time correlate positively with the distance between home and user, measured as the distance between home  and the phone. We write below this distance as the variable ``Home $\Leftrightarrow$ phone”. This distance is very important in this analysis, being home the location where the correctness of the user answers is computed. This correlation has a series of negative consequences on the answer quality, as shown in Fig. \ref{Reaction Time} \& \ref{Completion time}. In fact, a longer reaction time (Section \ref{sec:H3}, Fig. \ref{Chain of errors}), has two effects. The first is a direct effect on increasing the distance ``Home $\Leftrightarrow$ phone” (the more the user waits, the more s/he is moving away from home). The second indirect effect occurs through the increase of the ``pending notifications" (the more the user waits, the more unanswered questions accumulate). This effect, in turn, (Section \ref{H2}, Tab. \ref{H2a} \& Section \ref{sec:H3}, Fig. \ref{Chain of errors}) induces a shorter completion time which in turn should decrease the error, as from the previous analysis. This result does not contradicts our analysis, but it shows the twofold way in which an error can appear. With long completion times we may have 
memory errors. However, when the same activity is repeated many times (e.g., a student quickly and ``automatically" filling a sequence of four notifications with the same answer at the end of a two-hour class),  there is a high risk of reducing attention with an increased risk of typing errors. 


\subsection{Recommendations for future ESM studies}
\label{sec-fr3}

The key lesson learned from this work is that \textit{the reaction time is the crucial factor to be controlled in order to improve the answer quality}, see the discussion at the end of Section 4.3. The completion time is anyhow much harder to control, because of the many contradicting 
factors influencing it and also because of the small difference between the time taken to complete a correct answer and the time taken to complete an incorrect answer (see again the discussion at the end of Section 4.3).
In order to improve the reaction time, it is of paramount importance to have accurate information about where the user is, what he is doing and with whom, namely the current context in its various dimensions, e.g., situational, social, temporal. As it emerges from this model, one cannot simply note whether the phone is \textit{on/off} \cite{wang2018tracking} or whether the subject is changing activity \cite{H-2017-Pielot,S-1986-Holland}. It is the combination of the different contextual dimensions, combined with time, that defines with  greater precision when a user is most likely to respond, and to provide a correct answer.

There is evidence that, if we want to study human activities in the wild, probably we cannot limit our observation windows to a couple of weeks, as also suggested in \cite{wang2018tracking}, nor to a single historical moment. Human activities change according to the social time activities, e.g., weekdays and weekends, change according to the season, and change according to social relations. There is no single best practice for doing ESM research. This depends on (a) the research question and research design (e.g., event-based sampling; time-based sampling; combination of time- and events-base); (b) the phenomenon under investigation, its frequency and regularity of occurrence, complexity, and interaction with other factors that need to be controlled (e.g., social context); and (c) the duration of the survey. So, as with traditional survey research, we need to continue to evaluate and report the limits and our strategies. The goal is to limit errors and their consequences.

Based on these considerations, and also on the experience matured in the work described in this paper, we can provide the following recommendations about to concrete organize the various aspects of an EMA/ESM experiment.

\vspace{0.1cm}\noindent
(a) Avoid voluntary participation. As several research studies have shown, participants must be paid  \cite{aminikhanghahi2019context, S-2019-Keusch}. No payment increases user dropout and non-cooperation.

\vspace{0.1cm}\noindent
(b) Avoid short time limits for responding to notifications. Over time, the user finds his or her own response routine. On the one hand, a long reaction time reduces the workload; on the other hand, it can increase memory errors. However, it is better to have a larger number of responses, which can be possibly excluded from the analysis, than no information at all. 

\vspace{0.1cm}\noindent
(c) Avoid complex cognitive tasks and limit the questions to simple, factual information, especially in the case of a long observation period (weeks or more).

\vspace{0.1cm}\noindent
(d) Evaluate the data collection duration and the notification frequencies according to the topics under study. If habits or specific daily routines are studied, the observation time and notification frequencies should be designed on the temporal shape of the evolution of the phenomenon, e.g., days or weeks.

\vspace{0.1cm}\noindent
(e) Consider both short and too long completion times as part of the response set (in these situations, the user often gives the same answer). These answers are useful to compute  the median completion and the diversity of answers (for instance, when be compared with the other users).

\vspace{0.1cm}\noindent
(f) Use the smartphone sensors to evaluate the probability that an answer is correct. For example, in order to find the probability of being distracted, or to predict the reaction time.

\begin{figure}[htp]
\centering
\includegraphics[scale=0.95]{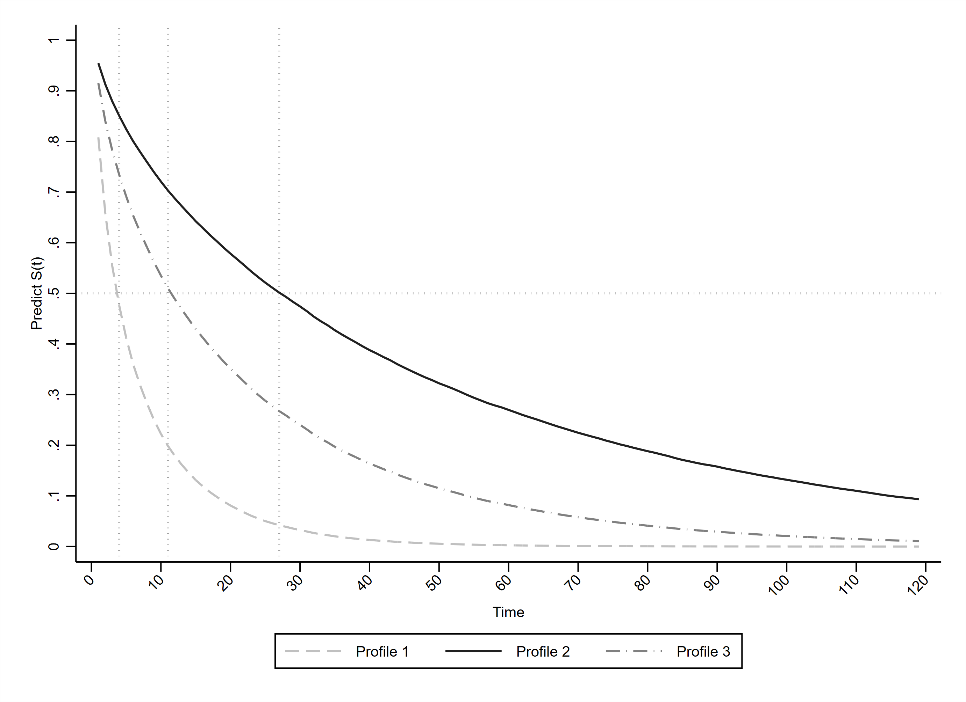}
\caption{Predicted survival function associated to three different user profiles.}
\label{user profile}
\end{figure}

\vspace{0.1cm}\noindent
(g) Prioritize sending notifications when the phone is in active use (better when the user is on social media), and when the user is alone (better in places where there is a high probability that the mood is lower or the subject is bored, e.g., travel or self-care). A good such example is Profile 1 of Fig. \ref{user profile}, where 
Fig. \ref{user profile} provides an estimated survival curve of the reaction time from three different participant profiles, as follows:
\begin{itemize}
\item 
\textbf{Profile 1}: (\textit{What}) Social media/chat – (\textit{With whom}) Alone – (\textit{Where}) University – (\textit{Mood}) sad - (\textit{Procrastination syndrome}) Low. The median reaction time is 4 min.; 
\item
\textbf{Profile 2}: (\textit{What}) Free time - (\textit{With whom}) Partner - (\textit{Where}) Home – (\textit{Mood}) Happy – (\textit{Procrastination syndrome}) High. The median reaction time is 27 min.; 
\item 
\textbf{Profile 3}: (\textit{What}) Study - (\textit{With whom}) Classmate - (\textit{Where}) University – (\textit{Mood}) neutral -\textit{Procrastination syndrome} (on average). The median reaction time is 11 min. 
\end{itemize}

\vspace{0.1cm}\noindent
(h) Finally, avoid as much as possible involving users with high procrastination syndrome. Look, for example, at Profile 2 in Fig. \ref{user profile}. When 50.0\% of Profile 2 users have filled out the notification, about 75.0\% of Profile 2 and 95.0\% of Profile 1 users have done the same.

\noindent
Two conclusive remarks. The first is that, assuming that the focus should be on how to improve the reaction time, then the next research question that needs to be answered is: \textbf{What is the best time to ask a question?} But this raises the question of how to do it. Our answer is that we should focus on minimizing the exogenous effects due to the context history. The future research will need to develop a holistic approach to the problem, so that \textit{systems will be able to learn what are right time}, \textit{to ask a question} taking into account i.e., the best situational contexts and, for what is possible, the subject's endogenous factors. In this perspective, some of them, e.g., procrastination syndrome and personality, can be computed once for all and can therefore be taken as input parameters to the machine learning algorithm.
The second remark is that the analysis provided in this paper is based on the data collected from a population of students. The selection of the sample is motivated by the fact students are easier to reach and also more prone to innovation and research experiments. It is not by chance that the choice made here  follows a long tradition of papers working on this population, see, e.g., \cite{wang2015smartgpa,wang2018tracking,zhang2021putting,giunchiglia2018mobile,ben2017crosscheck}.
The full generality of the results provided, future studies can, of course, be achieved by extending this type of study to other populations. 

%% file: section/8.Conclusion.tex
\section{Conclusion}
\label{sec:conclusion}

In this work, we have investigated the effects of various factors on the correctness of answers. Our study, based on an empirical analysis of a large dataset of daily behaviours, captures a rich and multifaceted picture of individual behaviours on potential interaction errors.

Our results suggest that while they are very useful in predicting certain contextual patterns, subjective annotations present a certain degree of error due to both exogenous and endogenous factors affecting the quality of responses. 
When focusing on research studies where the user is asked to provide data to a third party, these problems are in addition to many others which are already known. Some examples are: the social desirability effect that may prevent the study participant from reporting certain (socially disapproved) activities \cite{S-2003-Corbetta}; unreported activities when the participant perceives them as an intrusion on his or her privacy \cite{S-2015-Callegaro}; and the incorrect design of the data collection instrument, for example, the lack of an exhaustive list of response alternatives that are allowed to the respondent \cite{S-2019-Blaikie,S-2011-Groves}. Furthermore, in practical applications, collecting self-reported annotations is not always an option.